%% file: code.tex
\documentclass[aip,jcp,reprint,longbibliography]{revtex4-2}
\usepackage{amsmath}
\usepackage{graphicx}
\usepackage{tabularx}
\usepackage{hyperref}
\usepackage[utf8]{inputenc}
\usepackage[normalem]{ulem}
\usepackage{bbold}[frame=single]
\hypersetup{colorlinks=true, linkcolor=blue, citecolor=blue, urlcolor=blue}
\usepackage{fancyvrb}

\begin{document}
\title{ACE: A general-purpose non-Markovian open quantum systems simulation toolkit
based on process tensors}
\author{Moritz Cygorek}
\affiliation{Condensed Matter Theory, Department of Physics, TU Dortmund, 44221 Dortmund, Germany}
\affiliation{SUPA, Institute of Photonics and Quantum Sciences, Heriot-Watt University, Edinburgh EH14 4AS, United Kingdom}
\author{Erik M. Gauger}
\affiliation{SUPA, Institute of Photonics and Quantum Sciences, Heriot-Watt University, Edinburgh EH14 4AS, United Kingdom}

\begin{abstract}
We describe a general-purpose computational toolkit for simulating open quantum
systems, which provides numerically exact solutions for composites of
zero-dimensional quantum systems that may be strongly coupled to multiple, quite general 
non-Markovian environments. 
It is based on process tensor matrix product operators (PT-MPOs), 
which efficiently encapsulate environment influences. 
The code features implementations of several PT-MPO algorithms, in particular, 
Automated Compression of Environments (ACE) for general environments comprised 
of independent modes as well as schemes for generalized spin boson 
models. The latter includes a divide-and-conquer scheme for periodic PT-MPOs, 
which enable million time step simulations for realistic models.
PT-MPOs can be precalculated and reused for efficiently probing different 
time-dependent system Hamiltonians. They can also be stacked together and 
combined to provide numerically complete solutions of
small networks of open quantum systems.
The code is written in C++ and is fully controllable by configuration files,
for which we have developed a versatile and compact human-readable format.
\end{abstract}
\maketitle

\section{Introduction}
Many problems in quantum chemistry~\cite{Makri_ring,T-TEDOPA}, 
quantum optics~\cite{Carmichael1993,CygorekCoop}, 
condensed matter physics~\cite{Reiter_distinctive_characteristics}, 
and quantum information theory~\cite{PT_PRA} take the form
of (zero-dimensional) few-level open quantum systems coupled to some environment.
If the coupling is weak, standard perturbative and Born-Markov treatments
can be employed to derive time-local Lindblad master 
equations~\cite{BreuerPetruccione, Lindblad}, which are straightforward to solve 
numerically using standard differential equation algorithms~\cite{NumericalRecipes} or using convenient toolkits such as QuTiP~\cite{QuTiP} or QuantumOptics.jl~\cite{quantumopticsjl}.

The situation is more challenging when the system-environment coupling is strong
and non-Markovian memory effects have to be accounted for~\cite{RevModPhys_deVega}.
Then, an accurate treatment of environment effects requires 
modeling---explicitly or implicitly---the quantum dynamical evolution of the
environment. Because real environments typically consist of a (quasi)continuum of degrees of
freedom or modes, a many-body quantum systems arises, whose direct solution is
in general intractable. Most methods for non-Markovian open 
quantum systems tackle this challenge by focusing on a particular class 
of problems and make use of the particularities to reduce the problem 
complexity.

For example, if the environment is one-dimensional~\cite{PRX_Goold} or can 
be mapped onto one dimension~\cite{T-TEDOPA,Chin_noise-assisted}, 
tensor network structures like matrix product states (MPSs) and 
operators (MPOs)~\cite{MPS_Schollwoeck} provide very efficient 
numerically tractable representations of the state of the environment.
Larger-dimensional environments are often accurately modeled using 
mean-field or
cumulant expansion techniques~\cite{RevModPhys_Kuhn,DMS_nonmagnetic}
or treatments motivated by perturbation theory~\cite{Axt_DCT}.

One of the most widely studied classes of open quantum systems is the
spin-boson model, which describes (bio-)molecules~\cite{Makri_ring,T-TEDOPA}, 
resonant nanojunctions~\cite{junctions},
as well as semiconductor nanostructures like quantum dots (QDs)~\cite{Krummheuer}.
The particular Gaussian character of the linear coupling to a bath of harmonic 
oscillators enables a treatment using path integrals~\cite{FeynmanVernon},
which has been the basis of several practical methods. On the one hand, path integrals have been used to derive hierarchical equations of motion (HEOM)~\cite{HEOM89,HEOM90}, which are now a well established technique and implemented in several computer codes 
like QuTiP-BoFiN~\cite{Lambert23_bofin} and  HierarchicalEOM.jl~\cite{HEOMjl}.
On the other hand, there are schemes to sum up the path integral approximately or exactly in ways that avoids the exponential scaling of the Feynman-Vernon expression with respect to the number of time steps $n$. Examples are the iterative
path integral scheme QUAPI~\cite{QUAPI1,QUAPI2}, which reduces complexity by assuming a finite memory time. This method is only exponential in the number of time steps $n_\textrm{mem}$ within the memory time. The base of the exponential scaling can be reduced by the blip decomposition~\cite{Makri_Blip}, which also provides a very quickly converging approximate summation scheme for nearly incoherent dynamics. The Small Matrix Path Integrals (SMatPI)~\cite{SMatPI} decomposition using a series of small matrices of the dimension of the squared system Hilbert space strongly reduces the memory requirements and is implemented in the code PathSum~\cite{PathSum}.

Recently, tensor network representations have been exploited for efficient open quantum system simulations.
In the method TEMPO~\cite{TEMPO}, the augmented density matrix of QUAPI is represented as a MPO. While tensor network representations permit direct contraction schemes like PC-TNPI~\cite{Bose_TC-TNPI}, more commonly the key to their success lies in MPO compression~\cite{MPS_Schollwoeck}, which has an established role in leading to a very efficient representation of one-dimensional structures. This general principle has also been transferred to other open quantum systems techniques like ML-MCTDH~\cite{TN_ML-MCTDH} and HEOM~\cite{TN_HEOM}. 

\begin{figure}
\includegraphics[width=\linewidth]{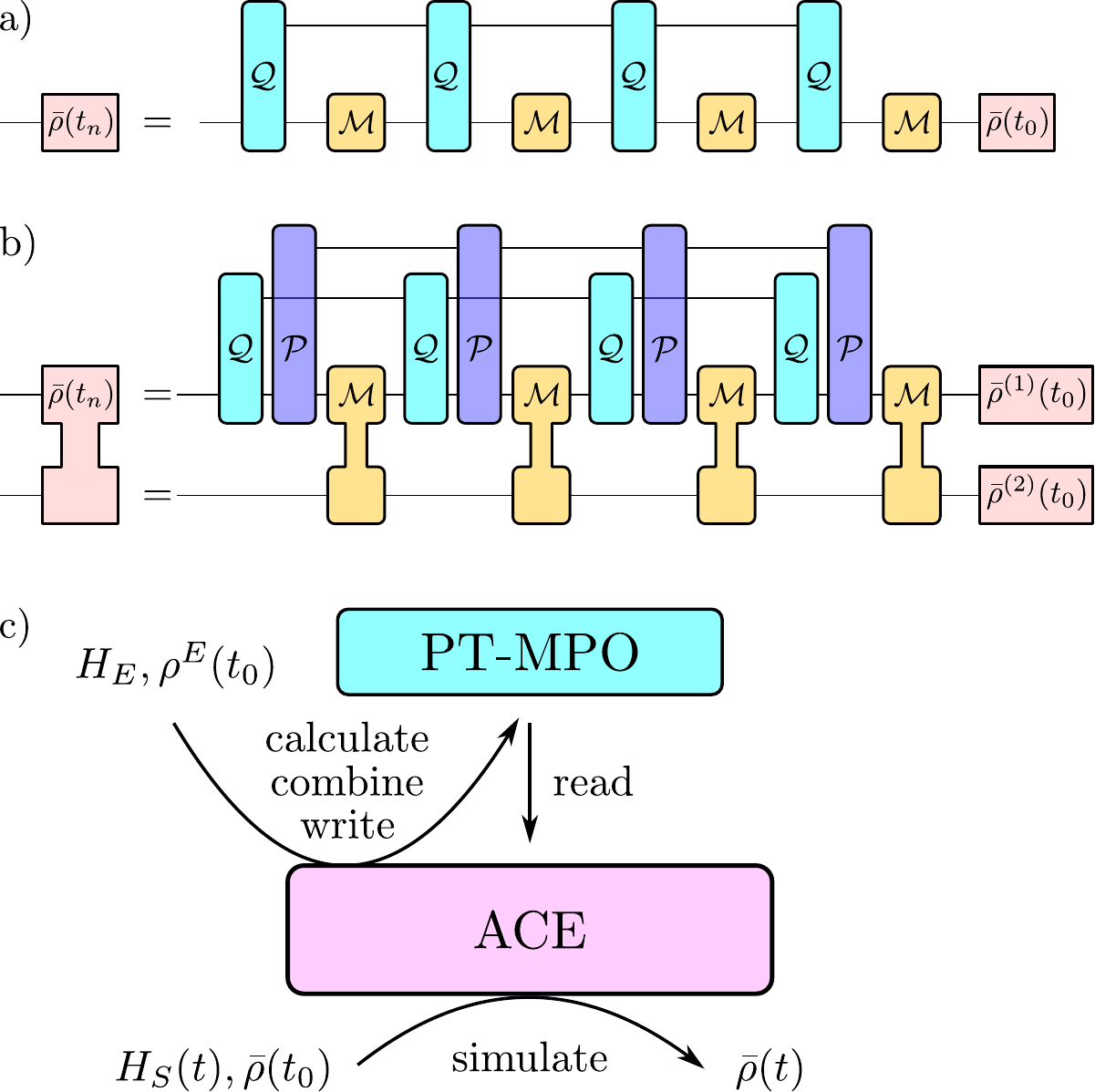}
\caption{\label{fig:sketch}(a): Propagation of a quantum system (reduced 
density matrix $\bar{\rho}$) coupled to an environment described by a 
PT-MPO (set of $\mathcal{Q}$'s) and subject to a free system propagator 
$\mathcal{M}$, which depends on the system Hamiltonian.
(b): Example for composition of open quantum systems. Depicted is a 
bipartite quantum system of interest, whose parts are coupled via the free
system propagator $\mathcal{M}$. The upper part of the system is coupled
to two (in general non-Markovian) environments, each described by a PT-MPO.
(c): Workflow of the ACE code. First, the ACE code can be used 
calculate a PT-MPO from the microscopic environment Hamiltonian and the 
initial state of the environment. 
Optionally the PT-MPO can be written to a file, read from file,
and combined with other PT-MPOs. 
Then, for a given system Hamiltonian and initial state, the reduced density
matrix and thereby all system observables can be extracted.
}
\end{figure}

At the core of our code is the process tensor (PT) formalism~\cite{PT_PRA, JP}, where environment
influences are encapsulated and represented in efficient tensor network 
structures called 
process tensor matrix product operators (PT-MPOs)~\cite{inner_bonds}.
PT-MPOs can be constructed to depend only on the environment Hamiltonian and the system-environment
interaction, and describe the impact of the environment irrespective of  
interventions performed on the system, such as unitary time evolution due
to a time-dependent system Hamiltonian or measurements~\cite{PT_PRA} [see Fig.~\ref{fig:sketch}(a)].
This has many advantages: First, the numerically challenging part, the PT-MPO
calculation, has to be performed only once and the resulting PT-MPO can be 
reused many times, e.g., to optimize parameters or driving protocols 
for open quantum systems~\cite{Fux_PRL}. 
Second, the allowed interventions on the system include those needed 
to extract multi-time correlation functions. This is particularly
useful for non-Markovian open quantum systems when the 
quantum regression theorem no longer holds~\cite{PI_QRT}.
Finally, a quantum system coupled to
two or more environments can be simulated using two PT-MPOs calculated 
independently of each other. The result remains numerically 
exact~\cite{ACE,inner_bonds}. This fact can be used to investigate
non-additive multi-environment effects~\cite{twobath,WiercinskiPolaronDressing} [see Fig.~\ref{fig:sketch}(b)]
as well as cooperative effects in multi-site quantum systems where each site is coupled 
to a local non-Markovian environment~\cite{CoopWiercinski}. 
Due to the modularity and separation of concerns, PT-MPOs are promising for 
scalable schemes to simulate small to medium-sized quantum 
networks~\cite{Fux_spinchain}.

The first algorithms to calculate PT-MPOs~\cite{JP} started from a 
tensor network derived from path integrals~\cite{TEMPO}, and were thus 
restricted to Gaussian environments like generalized spin-boson models.
These are now implemented in the Python package OQuPy~\cite{OQuPy}.
Subsequently, progress has been made in two directions: First, new schemes
provide orders-of-magnitude speed-up by employing a divide-and-conquer
strategy~\cite{DnC} and periodic PT-MPOs~\cite{DnC,Strunz_infinite}.
Second, algorithms for different~\cite{YeChan} and more general~\cite{ACE} types of environments have become available. Specifically, 
in Ref.~\cite{ACE}, we introduced the algorithm 
\textit{Automated Compression of Environments} (ACE), which is applicable to
any environment that can be described in terms of $N_E$ independent modes,
such as phonon, photon, fermion, spin, and anharmonic environments. 
Moreover, the environment modes themselves may be subject to Markovian losses
and they may be time-dependently driven. Changing the contraction order of
the tensor network and employing an efficient `preselection' scheme for
the combination of PT-MPOs with large inner bonds yields a variant of ACE
which is about one to two orders of magnitude faster~\cite{combine_tree}.

In this article, we describe our accompanying eponymous numerical toolkit ACE~\cite{ACEcode}, which implements the
ACE method as well as other PT-MPO techniques. It is designed to allow users to profit
from the efficiency, modularity, and generality of PT-MPO techniques without
requiring any programming. The physical problem is instead defined in 
human-readable configuration files, where one specifies the microscopic 
Hamiltonians, initial states, as well as a set of control and 
convergence parameters. For convenience, shortcut notations are available for some special and recurring problem classes. 
One can also easily switch between several algorithms and thus quickly
compare their performance and accuracy.

The article is structured as follows: In Sec.~\ref{sec:methods}, we summarize
the fundamentals of PT-MPOs as well as the implemented methods.
In Sec.~\ref{sec:code}, we describe the general usage of the ACE code, which is followed by a series of concrete examples in Sec.~\ref{sec:examples}. A summary of available commands for configuration files is provided in the appendix~\ref{app:commands}.

\section{Implemented methods~\label{sec:methods}}
\subsection{PT-MPOs: General principles}
The ACE code uses PT-MPOs to numerically exactly simulate non-Markovian
open quantum systems. We therefore first sketch the key concepts of 
PT-MPOs. A more detailed derivation can be found in 
Ref.~\cite{inner_bonds}.

The time evolution of a system (S) together with its environment (E) is
formally given by
\begin{align}
\rho(t) = & \overleftarrow{T} \exp\bigg({ \int\limits_0^t d\tau\, \mathcal{L}(\tau) }\bigg) \rho(0),
\end{align} 
where $ \overleftarrow{T} $ is the time ordering operator and $\mathcal{L}$
is the total Liouvillian, which determines the time evolution 
$\dot{\rho}(t)=\mathcal{L}(t) \rho(t)$, e.g.,
$\mathcal{L}\rho=-\frac i\hbar[H,\rho]$ with total Hamiltonian $H$.
The total Liouvillian 
\begin{align}
\mathcal{L}=&\mathcal{L}_S + \mathcal{L}_E
\end{align}
is split into a part $\mathcal{L}_S$, which only affects system degrees 
of freedom, and an environment part $\mathcal{L}_E$, which affects both,
the environment and the system, as it also includes the system-environment
interaction. 
Next, one introduces a time grid $t_j=t_0+j \Delta t$ 
with steps $\Delta t$, which is chosen
small enough to ensure (i) that $\mathcal{L}$ can be considered constant in 
time over the time step $\Delta t$ and (ii) that the error of the Trotter splitting 
\begin{align}
\label{eq:asymTrotter}
e^{\mathcal{L}\Delta t} =e^{\mathcal{L}_E\Delta t} e^{\mathcal{L}_S\Delta t}
+\mathcal{O}(\Delta t^2)
\end{align}
is small enough.
Tracing over the environment at the final time step $t_n=t$ yields
the reduced system density matrix $\bar{\rho}(t)=\textrm{Tr}_E\big\{\rho(t)\big\}$
\begin{align}
\label{eq:explicit_evolution}
\bar{\rho}(t)=&
\textrm{Tr}_E\Big\{e^{\mathcal{L}_E\Delta t} e^{\mathcal{L}_S\Delta t}
\dots e^{\mathcal{L}_E\Delta t} e^{\mathcal{L}_S\Delta t} 
\bar{\rho}(0)\otimes \rho^E(0)\Big\},
\end{align}
where we additionally assume that the total system initially factorizes
$\rho(0)=\bar{\rho}(0)\otimes \rho^E(0)$ in system and environment parts.
Collecting all terms relating to the environment, we rewrite 
Eq.~\eqref{eq:explicit_evolution} as
\begin{align}
\bar{\rho}_{\alpha_n}=&\sum_{\substack{
\alpha_{n-1},\ldots,\alpha_0 \\
\alpha'_n,\ldots,{\alpha}'_1}}
\mathcal{I}^{(\alpha_n,{\alpha}'_n)\ldots (\alpha_1,{\alpha}'_1)}
\bigg(\prod_{l=1}^n
\mathcal{M}^{\alpha'_l \alpha_{l-1}} \bigg)
\bar{\rho}_{\alpha_0},
\label{eq:feynman_PI}
\end{align}
where we have introduced the notation that left and right system indices
$\nu_j,\mu_j$ on the density matrix at time $t_j$ are combined to a single
Liouville space index $\alpha_j=(\nu_j,\mu_j)$ and the time argument is
implied in the sub-index, e.g, 
$\bar{\rho}_{\alpha_j}=\bar{\rho}_{\nu_j,\mu_j}(t_j)$.
Furthermore, $\mathcal{M}^{\alpha'_l \alpha_{l-1}} = 
\big(e^{\mathcal{L}_S\Delta t}\big)_{(\nu'_l,\mu'_l),(\nu_{l-1},\mu_{l-1})}$ 
is the explicit matrix representation of the system propagator 
$e^{\mathcal{L}_S\Delta t}$ and 
$\mathcal{I}^{(\alpha_n,{\alpha}'_n)\ldots (\alpha_1,{\alpha}'_1)}$
is the generalized\footnote{Feynman and Vernon assume in Ref.~\cite{FeynmanVernon} 
that the system of interest couples to the bath via the position coordinate.
This corresponds to a diagonalizable coupling as, e.g., in the spin-boson model,
but does not capture more general couplings via a sum of terms, where the
system operators cannot be simultaneously diagonalized. In the latter case,
the influence functional must be able to induce transitions between states,
which corresponds to entries with index combinations $(\alpha_l, \alpha'_l)$
for $\alpha_l\neq\alpha'_l$. If the coupling is diagonal, i.e. $\propto
\delta_{\alpha_l,\alpha'_l}$, then a single index per time step is sufficient,
and one obtains the original form of the Feynman-Vernon influence functional
in Ref.~\cite{FeynmanVernon}.
} Feynman-Vernon influence functional~\cite{FeynmanVernon}.

Eq.~\eqref{eq:feynman_PI} is universally valid and exact up to the (controllable) Trotter error, 
but it suffers from exponential scaling of the number of summands 
with the number of time steps. 
PT-MPOs address this by representing the generalized influence functional
in matrix product operator form
\begin{align}
&\mathcal{I}^{(\alpha_n,\alpha'_n)\ldots (\alpha_1,\alpha'_1)}
\nonumber\\ &\quad\quad\;=
\sum_{d_{n},\dots,d_0}
\mathcal{Q}^{(\alpha_{n},\alpha'_{n})}_{d_n d_{n-1}}
\mathcal{Q}^{(\alpha_{n-1},\alpha'_{n-1})}_{d_{n-1} d_{n-2}}
\ldots
\mathcal{Q}^{(\alpha_1,\alpha'_1)}_{d_1 d_0},
\label{eq:IfromQ}
\end{align}
which can be viewed as a series of matrix products with respect to inner bonds
$d_l$. On the edges, the inner bonds only take one value $d_n=d_0=0$.

Then, Eq.~\eqref{eq:feynman_PI} becomes 
\begin{align}
\bar{\rho}_{\alpha_n}=& \sum_{d_n} q_{d_n}
\bigg(\prod_{l=1}^n 
\sum_{\alpha'_{l},\alpha_{l-1},d_{l-1}}\!\!\!
\mathcal{Q}^{(\alpha_l,{\alpha}'_l)}_{d_l, d_{l-1}}
\mathcal{M}^{\alpha'_l \alpha_{l-1}} \bigg)
\bar{\rho}_{\alpha_0} \delta_{d_0,0},
\label{eq:iter}
\end{align}
which can be propagated time step by time step, and thus reduces the 
numerical complexity from exponential to polynomial in the number of time steps. 
The reduced system density matrix at intermediate time steps
can be obtained my contracting the inner bond $d_n$ with the closure $q_{d_n}$, which can be calculated
from the PT-MPO as discussed in Ref.~\cite{ACE}.
Equation~\eqref{eq:iter} is visualized in Fig.~\ref{fig:sketch}(a).

In principle, Eq.~\eqref{eq:iter} reproduces   Eq.~\eqref{eq:explicit_evolution} if one identifies the PT-MPO matrices $\mathcal{Q}$ with the environment propagators $e^{\mathcal{L}_E\Delta t}$. However, this choice would entail dimensions of inner bonds
of the size of the full environment Liouville space. This is numerically intractable,
especially for environments consisting of a continuum of modes. 
The advantange of MPO representations is that efficient compression schemes are available
that reduce the inner bond dimensions while conserving the action on the outer bonds.
This is achieved by sweeping across the MPOs while performing singular value decompositions
(SVDs) and keeping only large singular values $\sigma_k\ge \epsilon \sigma_0$, 
where $\sigma_0$ is the largest singular value and $\epsilon$ defines the threshold.

Thus, when the environment influences are represented as a two-dimensional tensor network,
this network can be sequentially contracted row by row to eventually yield a PT-MPO
describing the full influence. After each contraction step, line sweeps (forward and
backward) are performed to keep the bond dimensions tractable at all times. 
Different initial tensor networks are considered in different algorithms, and also
blockwise combination is considered as an alternative to sequential contraction.

In recent work~\cite{inner_bonds} we showed that the PT-MPO matrices $\mathcal{Q}$ 
can be viewed as the environment propagator $e^{\mathcal{L}_E\Delta t}$ projected onto a subspace
of environment excitations $\mathcal{Q} = \mathcal{T} e^{\mathcal{L}_E\Delta t} \mathcal{T}^{-1}$ 
with lossy compression matrices $\mathcal{T}$ and
their pseudo-inverses $\mathcal{T}^{-1}$. The role of MPO compression is that 
it leads to the automatic selection of the most relevant subspace required for
an accurate description of the systems dynamics.

This concept also explains why one obtains numerically exact solutions for 
systems coupled to multiple environments when the corresponding PT-MPOs
are stacked together as depicted in Fig.~\ref{fig:sketch}(b). The same 
panel further shows how composite quantum systems can be propagated with 
PT-MPOs that have been calculated assuming coupling of only one part of 
the system to the environment. This provides the basis for numerically
complete simulations for small networks of open quantum systems~\cite{Fux_spinchain,CoopWiercinski}.

With the methodological background clarified, the workflow of the
ACE code, which is depicted in Fig.~\ref{fig:sketch}(c), is now easy to understand:
Before a non-Markovian open quantum system can be simulated, one first
has to obtain the corresponding PT-MPO(s). Several algorithms (see below) 
to this end are implemented. PT-MPOs can be calculated on the fly, i.e. kept
only in working memory, and used for a single simulation run. Alternatively,
it can be written to a file and reused for many simulations with different
system Hamiltonians, e.g. for optimizing system parameters, 
for identifying optimal driving protocols~\cite{Fux_PRL}, or for combining multiple
PT-MPOs in multi-environment simulations~\cite{CoopWiercinski}.
Eventually, the open quantum system is propagated using Eq.~\eqref{eq:iter}.

The following schemes for calculating PT-MPOs are implemented:
\subsection{Automated Compression of Environments}
The ACE algorithm~\cite{ACE} is extremely general. It can be applied to general
environments that consist of $N_E$ independent modes. Consequently, the environment
Liouvillian can be decomposed as
\begin{align}
\mathcal{L}_E=\sum_{k=1}^{N_E} \mathcal{L}_E^{(k)},
\end{align}
where $\mathcal{L}_E^{(k)}$ only affect the system and the $k$-th environment
mode. Similarly, the initial states of the modes are uncorrelated 
$\rho^E(t_0)=\prod_{k=1}^{N_E}\rho^{E,(k)}(t_0)$.
Then, PT-MPOs are calculated for each environment mode independently 
by identifying the PT-MPO matrices $\mathcal{Q}$ with the propagators 
$e^{\mathcal{L}_E^{(k)}\Delta t}$, multiplying with $\rho^{E,(k)}(t_0)$ in
the first step and taking the trace in the last step~\cite{ACE,inner_bonds}.
Then, the PT-MPOs for the individual environment modes are combined together
one after the other, while after each combination the joint PT-MPO is 
compressed using sweeps with truncated SVDs. Eventually, one ends up 
with a PT-MPO containing the influences of all modes.

In Ref.~\cite{combine_tree}, we demonstrated a variant of ACE which is 
typically one to two orders of magnitude faster than the original ACE algorithm.
While in the original ACE algorithm the modes are sequentially incorporated 
into a single growing PT-MPO, one can instead combine PT-MPOs corresponding
to neighboring modes pairwise. The resulting PT-MPOs are again combined pairwise, 
so an overall ordering of the form of a binary tree emerges~\cite{combine_tree}.
The speed-up arises from the fact that most PT-MPO combination steps involve
smaller inner bond dimensions. However, the last few combination steps 
involve PT-MPOs with large dimensions, for which the usual
compression schemes~\cite{MPS_Schollwoeck} would be prohibitively demanding. 
This can be addressed by employing a preselection step based on SVDs 
of the individual PT-MPOs that are combined~\cite{DnC}.
The massive reduction of computation times come at the cost of increased
error accumulation, which however can be counteracted by fine-tuning
convergence parameters. This fine-tuning is discussed in Sec.~\ref{sec:fine_tuning}. 

\subsection{PT-MPOs for the spin-boson model}
One of the most frequently studied open quantum systems models is the 
(generalized) spin-boson model defined by the environment Hamiltonian
\begin{align}
\label{eq:spinboson}
H_E=\sum_k \hbar\omega_k b^\dagger_k b_k 
+\sum_k \hbar (g^*_k b^\dagger_k  + g_k b_k) \hat{A} + \Delta H_{PS},
\end{align}
where $b^\dagger_k$ and $b_k$ are boson creation an annihilation operators, $\omega_k$ is
the frequency of mode $k$, and $g_k$ is the corresponding coupling constant.
The general Hermitian operator $\hat{A}$ acts only on the system
Hilbert space. The term $\Delta H_{PS}=\sum_{k}(|g_k|^2/\omega_k)\hat{A}^2$
is usually added to subtract the polaron shift, i.e. absorb the energy renormalization
caused by the system-environment interaction into a redefinition of system energies.

If the initial state of the environment is thermal with temperature $T$, 
the spin-boson model is completely defined by the operator $\hat{A}$ and
the spectral density \mbox{$J(\omega)=\sum_k |g_k|^2 \delta(\omega-\omega_k)$}.  
In particular, the bosonic environment has Gaussian statistics, i.e. all environment
correlation functions can be reduced to the two-time correlation function
$C(t)=\sum_k |g_k^2| \langle b^\dagger_k(t) b_k(0)\rangle$, which can be
expressed as
\begin{align}
C(t) 
=&\int\limits_0^\infty d\omega\, J(\omega)\big[\coth(\beta\hbar\omega/2)\cos(\omega t)-i\sin(\omega t)\big].
\end{align}

The Gaussian character of the spin-boson environment facilitates the derivation of an
explicit expression of the Feynman-Vernon influence functional via 
path integrals~\cite{FeynmanVernon}. This has been used in the iterative path
integral method QUAPI~\cite{QUAPI1,QUAPI2}, which relies on the fact that the memory
of the environment, i.e. the support of the bath correlation function, is often finite
and contained within a few ($n_c$) timesteps. To combat the exponential scaling
of QUAPI with the number of memory timesteps $n_\textrm{mem}$,  Ref.~\cite{TEMPO} cast the QUAPI approach 
into a matrix product operator form, yielding the TEMPO algorithm. There the influence functional
for a generalized spin-boson model was represented as a tensor network.
Shortly after, J{\o}rgensen and Pollock~\cite{JP} realized that the same tensor network
representation of the influence functional that also appears in TEMPO  
can be contracted to yield a PT-MPO. Thereby they derived the first and currently most
commonly used PT-MPO method, which is implemented in the ACE code and also, e.g.,
in the OQuPy code~\cite{OQuPy}.

Recently~\cite{DnC}, we developed a divide-and-conquer scheme to contract the
tensor network for Gaussian environments. 
While the approach by J{\o}rgensen and Pollock~\cite{JP} requires
$\mathcal{O}(n^2)$ SVDs without memory truncation and 
$\mathcal{O}(n n_\textrm{mem})$ SVDs with memory truncation, the divide-and-conquer 
scheme is quasi-linear $\mathcal{O}(n \log n)$ if no memory truncation is used.
Moreover, if the memory can be truncated after $n_\textrm{mem}$ steps, it is possible
to calculate a periodically repeating block of PT-MPO matrices with
$\mathcal{O}(n_\textrm{mem}\log n_\textrm{mem})$ SVDs, which is independent of the total propagation
time $n$. However, these methods require the preselection approach for
combining PT-MPOs with large inner dimensions and hence may need fine-tuning
of convergence parameters for optimal results (see Sec.~\ref{sec:fine_tuning}).
Using them, solution of multi-scale problems involving propagation over
millions of time steps have been demonstrated in Ref.~\cite{DnC}. 

All of the above variants to calculate PT-MPOs for Gaussian environments
as well as the conventional QUAPI and original TEMPO algorithms are included and available within our ACE code. By contrast, 
the recent method by Link, Tu, and Strunz~\cite{Strunz_infinite} for PT-MPOs
consisting of a single repeating block has not yet been implemented.

\subsection{\label{sec:outer_reduction}Outer bond reduction}
An important aspect that affects the performance of PT-MPO techinques is the scaling 
with the dimension $D$ of the system Hilbert space. The outer bonds of
the PT-MPO matrices $\mathcal{Q}^{(\alpha_l,\alpha'_l)}_{d_l,d_{l-1}}$,
namely the set $\beta_l:=(\alpha_l,\alpha'_l)$, spans $D^4$ entries.
Thus, a naive implementation keeping all of these entries explicitly 
restricts PT-MPO techniques to very small system size. In particular, 
the $\mathcal{O}(D^4)$ scaling is then a major obstacle for `larger' systems because the total system dimension $D$ of composite systems scales
exponentially with the number of its parts.

Our strategy to deal with large outer bonds is to use the fact that in many 
situations there are several values of $\beta_l$ where 
$\mathcal{Q}^{\beta_l}_{d_l,d_{l-1}}=0$ for all combinations of $d_l$ and $d_{l-1}$, 
or where several  $\mathcal{Q}^{\beta_l}_{d_l,d_{l-1}}=\mathcal{Q}^{\beta'_l}_{d_l,d_{l-1}}$
for different $\beta_l\neq\beta'_l$. An example of the former is the case of the 
spin-boson model when the system coupling operator $\hat{A}$ in Eq.~\eqref{eq:spinboson}
is diagonal. Then, the PT-MPO matrices cannot directly induce transitions between system
states and
$\mathcal{Q}^{(\alpha_l,\alpha'_l)}_{d_l,d_{l-1}}\propto \delta_{\alpha_l,\alpha'_l}$, which
reduces the number of values of the outer bond indices to at most $D^2$. 
The situation where PT-MPO matrices with different values of $\beta_l$ are identical has 
also been discussed for QUAPI simulations in Ref.~\cite{PI_cavityfeeding}.
Translating these dicussions to PT-MPO techiques corresponds to utilizing degeneracies 
of eigenvalues of the spin-boson system coupling operator $\hat{A}$. These degeneracies
arise trivially when the environment is coupled only to one subsystem of a composite
open quantum system, e.g., for quantum dots coupled to an optical microcavity as well
as to a non-Markovian phonon bath~\cite{PI_singlephoton,PI_Fock,PI_cats}.
Moreover, degeneracies appear when there are decoherence-free subspaces, e.g. 
in the case of the biexciton-exciton diamond-shaped four-level system in a
quantum dot, where the two excitonic states with different spin selection rules
couple identically to the local phonon bath~\cite{PI_entangled_PRL,PI_Concurrence}.

In the ACE code, we therefore only store and operate on a single non-zero representation
of $\mathcal{Q}^{\beta_l}_{d_l,d_{l-1}}=0$, where $\beta_l$ is viewed as a dictionary
mapping the combination of physical indices $(\alpha_l, \alpha'_l)$ to matrices 
(with respect to $d_l$ and $d_{l-1}$) $\mathcal{Q}^{\beta_l}_{d_l,d_{l-1}}$.
In particular for the spin-boson model, the Hermitian coupling operator $\hat{A}$ is 
first diagonalized, the eigenvalues are checked for degeneracies, the PT-MPO is 
calculated for the reduced set of outer bonds, and---if $\hat{A}$ was not diagonal 
from the start---the outer bonds are expanded and rotated back to the original frame
undoing the diagonalization.
Moreover, the code provides the option to expand the outer bond dimensions temporarily
to facilitate the simulation of a composite open quantum system when the PT-MPO 
was calculated only accounting for the concrete subsystem the environment is
coupled to directly. This is key for making PT-MPO methods tractable for larger
multi-level systems as well as for small quantum networks.

\section{ACE Code\label{sec:code}}
\subsection{General structure and usage}
The ACE code is written in C++11 and can be fully controlled by configuration files.
Thus, it only has to be compiled once, and no C++ programming skills are required for
operation. The ACE code is freely and publicly available in Ref.~\cite{ACEcode}.

The only system requirement is that the header files of Eigen~\cite{Eigen} are present. 
The code can optionally be linked against LAPACK, which we find to be highly advantageous, 
especially when the implementation by Intel MKL is used.
A Makefile is available to facilitate compilation on Linux operating systems. 
Compilation has been tested on the Windows Subsystem for Linux and on macOS as well.

The code itself is composed of a library, whose functions are called by 
several binaries. In addition to the main binary \texttt{ACE} we provide a set 
of tools, e.g., to analyze or modify PT-MPOs. 
The binaries are controlled by command line options and/or configuration files.
For example, running  \texttt{ACE example.param -dt 0.01} from command line instructs
the code to process the file \texttt{example.param} 
(optional first argument; file name must not begin with a dash) 
and override the parameter \texttt{dt}, which describes the 
width of the time step, with the value 0.01. Alternatively, the time step could be 
specified in the configuration file \texttt{example.param} by adding a line \texttt{dt 0.01} 
(without the dash used for command line arguments). Any number of white spaces between 
parameter name and value are allowed.
These conventions facilitate running and processing a series of simulations with 
scripting languages including (Bash) shell scripts, PERL, and Python.

Before we explain the usage of the code on various examples, we
cover two general aspects: Parameters given as
\textit{matrix-valued expressions} and the format of input and output files.

\subsection{\label{sec:matrix-valued}Matrix-valued expressions}
The broad scope of ACE entails that a flexible way to specify Hamiltonians, 
initial states, observables, and other matrix-valued inputs is needed.
To this end, we developed a versatile notation for specifying matrix-valued 
expressions as text in input files or as command line arguments, which can
still be parsed with reasonable effort by the C++ program.
Our notation is inspired by standard mathematical notation for problems 
in quantum optics, Dirac's bra-ket notation, and second quantization.

\begin{table}
\renewcommand{\arraystretch}{1.25}
\begin{tabularx}{\linewidth}{|l| X|}
\hline
Expression & Value \\
\hline
\texttt{+}, \texttt{-}, \texttt{*}, \texttt{/} & Basic mathematical operations. Use \texttt{*} also for  matrix-matrix multiplications. \\
\texttt{(...)} & Parentheses\\
\texttt{pi} & $\pi=3.1415...$ \\
\texttt{hbar} & reduced Planck constant $\hbar=0.658..$ meVps\\
\texttt{kB} & Boltzmann constant $k_B=0.0861..$ meV/K \\
\texttt{wn} & Translation factor from wavenumbers to angular frequencies 
\mbox{0.188... cm/ps} \\
\texttt{sqrt(...)} & Square root function $\sqrt{...}$ \\
\texttt{exp(...)} & Exponential function $e^{...}$\\
\texttt{otimes} & Kronecker product $\otimes$\\
\texttt{|i><j|\_D} & Dirac operator $|i\rangle\langle j|$ on a $D$-dimensional Hilbert space; $i,j\in \{0,1,\dots,D-1\}$\\
\texttt{Id\_D} & $D$-dimensional identity matrix \\
\texttt{sigma\_x} &Pauli matrix $\sigma_x$ \\
\texttt{sigma\_y} &Pauli matrix $\sigma_y$ \\
\texttt{sigma\_z} &Pauli matrix $\sigma_z$ \\
\texttt{bdagger\_D}, \texttt{b\_D} & Bosonic creation and annihilation operators
truncated at Hilbert space dimension $D$\\
\texttt{n\_D} & Bosonic occupation number \texttt{bdagger\_D*b\_D} \\
\hline
\end{tabularx}
\caption{\label{tab:constants}Interpreted terms in matrix-valued expressions.}
\end{table}

Matrix-valued expressions are enclosed in curly braces. For example
\texttt{\{|0><1|\_2\}} represents the Dirac operator $|0\rangle\langle 1|$, 
which transfers the excitation from the excited state $|1\rangle$ to a ground state $|0\rangle$, in a two-level
system. These operators can be scaled and added as in 
\texttt{\{hbar/2*(|0><1|\_2 + |1><0|\_2)\}}, which represents the spin operator 
$\frac{\hbar}{2} \sigma_x$, where $\sigma_x$ is the usual Pauli matrix.
Some constants like \texttt{pi}=$\pi$ and \texttt{hbar}=$\hbar$ (meV ps) as well as functions and matrices are
prefined. These are listed in Tab.~\ref{tab:constants}.

The composition of operators acting on two subsystems or on a system and its environment
is facilitated by \texttt{otimes}. For example, the interaction part of the 
Jaynes-Cummings Hamiltonian with a 5-dimensional bosonic Hilbert space is written as
\texttt{\{|0><1|\_2 otimes bdagger\_5+|1><0|\_2 otimes b\_5\}}. 

The default units are assumed to be ps for time units, ps$^{-1}$ for rates and frequencies,
meV for energy units, and K for temperatures. These are suitable units for many platforms for
quantum technologies like solid state quantum emitters or molecules. 
Simulations for dimensionless problems are realized by multiplying all energy parameters with \texttt{hbar} and temperatures with
\texttt{hbar/kB}.  

Parameters consisting of single floating point numbers can be specified
as matrix-valued expressions for a 1x1 matrix, from which the real part is extracted. 
For example, one can set the final time of a simulation to $2\pi$ by specifying
\texttt{te \{2*pi\}}.

Finally, it should be noted that providing matrix-valued expression via the command 
line may require putting the expression additionally in double quotes, e.g., to prevent
the shell from parsing symbols like less \texttt{<} and greater \texttt{>} symbols.
The validity of a matrix-valued expression can be checked on the command line 
using the binary \texttt{readexpression} followed by an expression in double quotes and
curly braces.

\subsection{Input and output files}

Some parts of the problem specification may be described by functions, such as
pulse envelopes, spectral densities of environments, etc. Moreover, the simulation 
results---operator averages as a function of time---are stored in files.
For both, input and output, we use whitespace-separated plain text files organized in 
columns of floating point numbers in standard C/C++ format.
Any content after the symbol \texttt{\#} is regarded as a comment and thus ignored.
This format allows the data to be displayed directly with gnuplot~\cite{gnuplot}.
\begin{figure*}
\includegraphics[width=0.95\linewidth]{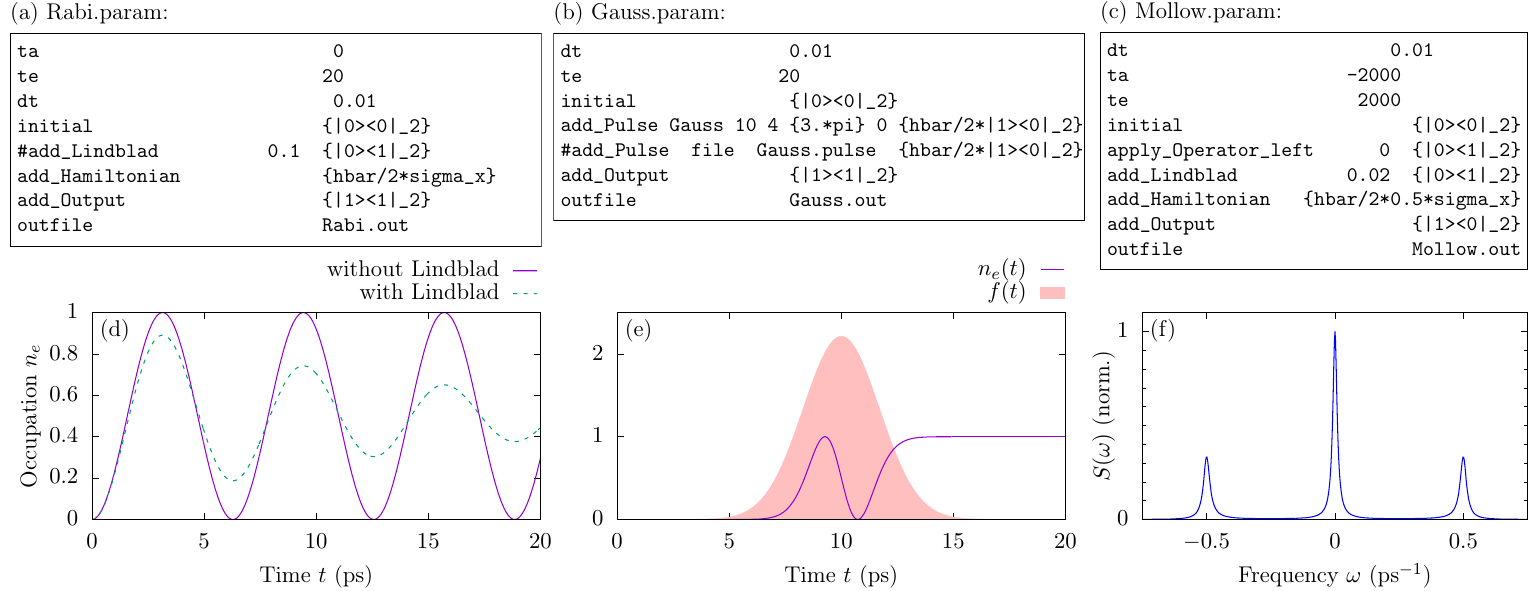}
\caption{\label{fig:Markov}(a)-(c): Configuration files and
(d)-(f): simulation results for examples of closed and 
Markovian quantum systems.
(a) and (d): Rabi oscillations of a continuously driven two-level system
with and without Lindbladian damping.
(b) and (e): Two-level system driven with a Gaussian $3\pi$ pulse.
(c) and (f): Extraction of first-order coherence and its 
Fourier transform for a driven two-level system, giving rise to a
Mollow triplet in the emission spectrum.}
\end{figure*}

For example, files containing pulse envelopes $f(t)$ are expected to contain 
times $t_j$ in the first column and real and imaginary parts of $f(t_j)$ 
in the second and third column, respectively.
Files for spectral densities $J(\omega)$, which describe how strongly Gaussian baths are coupled 
to the system at a given frequency $\omega$, have to contain two columns: the first
containing frequencies $\omega_j$ and the second containing the (real) value of
$J(\omega_j)$, both in ps$^{-1}$.

Output files contain time points $t_j$ in the first column. For each observable
specified using \texttt{add\_Output}, the operator average 
$\langle A(t)\rangle =\textrm{Tr}(A \rho(t))$ is extracted and represented as two
columns in the output file, corresponding to real and imaginary parts, respectively.
Due to their size PT-MPOs are stored as binary files and may be split into several
files each containing a block of at most $B$ PT-MPO matrices, where 
$B$ is specified by \texttt{buffer\_blocksize}.

\section{Examples\label{sec:examples}}
\subsection{Closed and Markovian quantum systems}
We begin with a simple example for the usage of the ACE code, starting with 
a continuously driven closed two-level system without any environment.
To this end, we run \texttt{ACE Rabi.param} with the configuration file
shown in Fig.~\ref{fig:Markov}(a).
There, the time grid is set to go from time \texttt{ta}=0 to \texttt{te}=20 ps 
in steps of \texttt{dt}=0.01 ps. 
The initial state (parameter \texttt{initial}) is set to the ground state 
$\bar{\rho}(0)=|0\rangle\langle 0|$ of a two-level system, which is specified as a 
matrix-valued expression.
We consider resonant Rabi driving using the Hamiltonian 
$H_S=\frac{\hbar}{2} (1\textrm{ ps}^{-1})\sigma_x$, where we use
the constant \texttt{hbar} to convert frequencies in ps$^{-1}$ to energies in meV.
The observable we are interested in is the occupation
$n_e(t)=\textrm{Tr}( |1\rangle\langle 1| \bar{\rho}(t))$.
This is specified using \texttt{add\_Output}, and the \texttt{outfile}
parameter instructs to code to save this observable in the output file \texttt{Rabi.out}.

Optionally, we add radiative loss modeled by the Lindbladian 
$(0.1\textrm{ ps}^{-1})\mathcal{D}_{|0\rangle\langle 1|}[\bar{\rho}]$ with
\begin{align}
\mathcal{D}_{\hat{A}}[\bar{\rho}]=
\hat{A}\bar{\rho}\hat{A}^\dagger - \frac12 (\hat{A}^\dagger\hat{A}\bar{\rho}
+\bar{\rho}\hat{A}^\dagger\hat{A}).
\end{align}
This term is commented out in the configuration file in Fig.~\ref{fig:Markov}(a), as all characters
in a line after the symbol \texttt{\#} are ignored. Thus, removing this symbol
we simulate the corresponding Markovian open quantum system.
The results with and without Lindblad term are depicted 
in Fig.~\ref{fig:Markov}(d) and show the expected (damped) Rabi oscillations.

More generally, quantum systems can be driven with pulses, i.e., 
with time-dependent system Hamiltonians $H_S(t)$.
In Fig.~\ref{fig:Markov}(b) and (e), we demonstrate the simulation of
a two-level system driven with a strong Gaussian pulse. Concretely, we apply
a system Hamiltonian 
\begin{align}
H_S(t)=(f(t)\hat{d} + f^*(t)\hat{d}^\dagger)
\end{align}
with a scalar function $f(t)$ and an operator $\hat{d}$ acting on the system
Hilbert space. We choose a Gaussian pulse
\begin{align}
f(t)=&\frac{A}{\sqrt{2\pi}\sigma}e^{-\frac 12\frac{(t-t_c)^2}{\sigma^2}}
e^{-i(\delta/\hbar)t},
\end{align}
with pulse area $A=3\pi$, pulse center $t_c=10$ ps, standard deviation 
$\sigma=\tau_\textrm{FWHM}/(2\sqrt{2\ln 2})$ with pulse duration 
$\tau_\textrm{FWHM}=4$ ps, and detuning $\delta=0$ meV.
Furthermore, we use the coupling operator  
$\hat{d}=(\hbar/2)|1\rangle\langle 0|$.

The time-dependent driving can be specified as indicated in 
Fig.~\ref{fig:Markov}(b), in one of two ways.
Either one can use a predefined \texttt{add\_Pulse} command.
Here, \texttt{Gauss} takes parameters in the following order:
$t_c$, $\tau_\textrm{FWHM}$, $A$, $\delta$, and $\hat{d}$.
Alternatively, as in the out-commented line in Fig.~\ref{fig:Markov}(b),
a pulse can be read from a file (here: file name \texttt{Gauss.pulse}) 
with three columns:
time points $t_j$ and real and imaginary parts of $f(t_j)$. 
The Hermitian conjugate part is added automatically.
This makes it possible to create completely arbitrary pulse shapes.

Finally, sometimes not only the density matrix is of interest but also 
multi-time correlation functions. For example, the emission spectrum
of a two-level system is related to the first-order coherences 
\mbox{$g^{(1)}(t,\tau)=\langle \sigma^+(t+\tau)\sigma^-(t) \rangle$}
by
\begin{align}
\label{eq:def_spectra}
S(\omega)=\textrm{Re}\int\limits_0^\infty d\tau\, 
\big[g^{(1)}(t,\tau)-g^{(1)}(t,\infty)\big] e^{-i\omega\tau}.
\end{align}
To evaluate the first-order coherences, we have to first propagate 
the system until it reaches a stationary state at time $t$, 
then apply the operator $\sigma^-=|0\rangle\langle 1|$ and propagate for
another time $\tau$, where the observable $\sigma^+=|1\rangle\langle 0|$ 
is extracted.
In Fig.~\ref{fig:Markov}(c), we depict the configuration file for
simulating a continuously driven two-level system from time 
\texttt{ta}=-2000 ps to time \texttt{te}=2000 ps.
At time $t=0$, the operator $|0\rangle\langle 1|$ is applied from the left
onto the density matrix, which is instructed with the command
\texttt{apply\_Operator\_left}, whose first argument is the time of application
and the second argument is the operator to be applied.
In Fig.~\ref{fig:Markov}(f), we present $S(\omega)$, which we obtain by 
Fourier transforming the second and third columns (real and imaginary parts) of the output file 
after cutting off all data in the output file prior to time $t$ and subtracting
the stationary part according to Eq.~\eqref{eq:def_spectra}.
A Mollow triplet is observed with the side peaks at the Rabi frequency
$\Omega=0.5$ ps$^{-1}$.

\subsection{Usage of the ACE algorithm}
\begin{figure*}
\includegraphics[width=\textwidth]{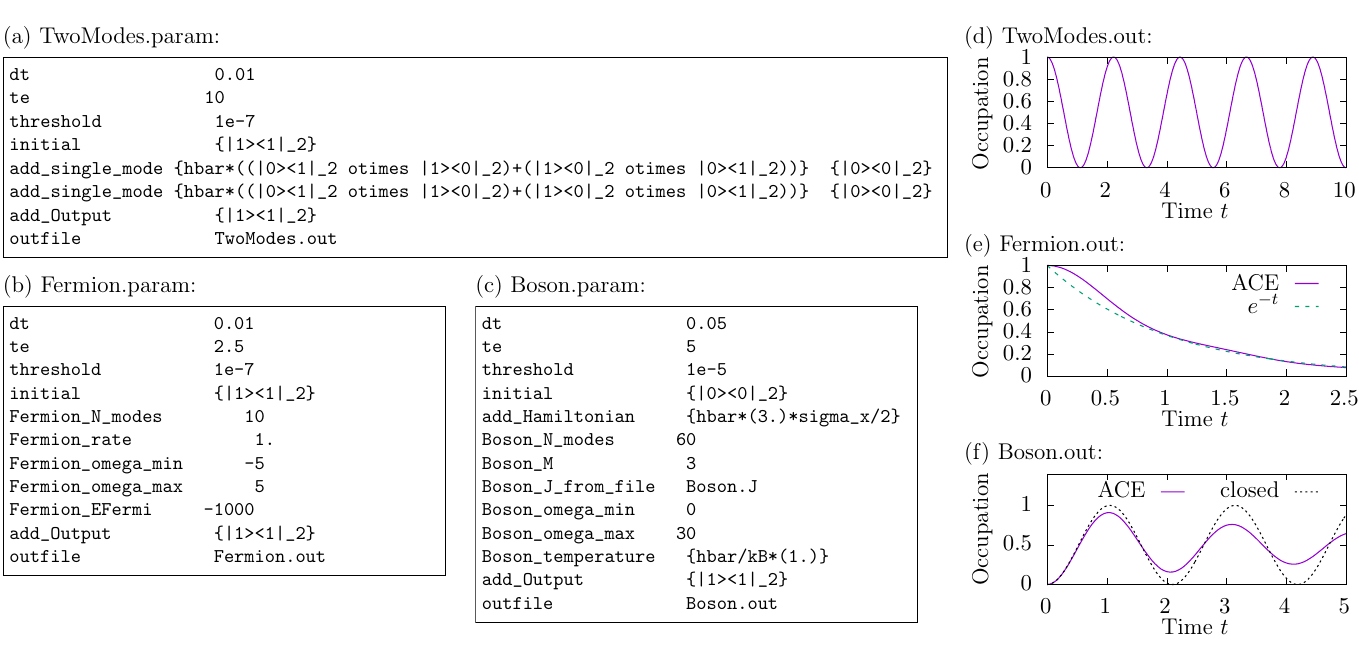}
\caption{\label{fig:example_ACE}Configuration files [(a),(b), and (c)] 
and results [(d), (e), and (f)] of examples using ACE: 
(a) and (d): Two-level system coupled to two environment modes, 
which are specified explicitly.
(b) and (e): Resonant level model with $N_E=10$ fermionic environment modes with
Markov approximation $e^{-t}$ for reference.
(c) and (f): Driven two-level system with and without 
spin-boson environment. The Ohmic spectral density is specified in
file \texttt{Boson.J}.}
\end{figure*}

Having described how closed and Markovian quantum systems can be simulated
using the ACE code, we now turn to non-Markovian open quantum systems.
As shown in Fig.~\ref{fig:example_ACE}, there are several ways to use the
ACE algorithm~\cite{ACE} to construct PT-MPOs from 
the explicit microscopic Hamiltonians of a set of $N_E$ environment modes and
to obtain the exact system dynamics up to an MPO compression error controlled by the 
\texttt{threshold} $\epsilon$ and Trotter error controlled by \texttt{dt}.

The first is to specify individual modes with the \texttt{add\_single\_mode} command,
as depicted in Fig.~\ref{fig:example_ACE}(a). The two arguments are
matrix-valued expressions for the mode Hamiltonian and for the  
initial state of the mode. Note that the mode Hamiltonian includes the coupling
to the system and thus acts on the space 
$\mathcal{H}_S\otimes\mathcal{H}_E^{(k)}$, while the inital state is a 
density matrix acting on the mode Hilbert space $\mathcal{H}_E^{(k)}$ alone.
In the example in Fig.~\ref{fig:example_ACE}(a), we couple a two-level
system to two identical modes via Hamiltonians
$H_E^{(k)}=\hbar g(\sigma^- c^\dagger_k + \sigma^+ c_k)$, where
$\sigma^\pm$ excite and destroy excitations in the central two-level system
and $c^\dagger_k$ and $c_k$ do the same for the $k$-th environment mode, which
is also a two-level system. The coupling constant is set to $g=1$ 
(technically in ps$^{-1}$, but identified with a dimensionless value). 
The central two-level system is initially occupied
while the environment modes are initially empty.
The result is depicted in Fig.~\ref{fig:example_ACE}(d), where one observes 
coherent oscillations between the central two-level systems and the symmetric
linear combination of the two environment modes.

Alternatively, for some frequently occuring environment models, 
the ACE code offers generators, which facilitate the convenient 
specification of a quasi-continuum of environment modes.
In Fig.~\ref{fig:example_ACE}(b) and (c), we use the \texttt{Fermion} and
\texttt{Boson} generators, respectively. Both require a series of similar 
commands starting with \texttt{Fermion\_} and \texttt{Boson\_}, respectively.
For example, \texttt{...N\_modes} defines the number of modes used for 
discretizing the continuum on a frequency interval defined by the limits
\texttt{...omega\_min} and \texttt{...omega\_max}. In the bosonic case in 
Fig.~\ref{fig:example_ACE}(c), the parameter \texttt{Boson\_M} determines the
size of the truncated Hilbert space per bosonic mode. For both types of 
environments, the \texttt{...temperature} can be set by the corresponding command.
If not set explicitly, a default value of $T=0$ is taken. The specified value
is assumed be given in units of Kelvin. An effectively dimensionless 
specification (typically denoted by $1/\beta$ in statistical physics) 
is achieved by mapping all energies to units of ps$^{-1}$, which we 
do in Fig.~\ref{fig:example_ACE}(c) by multiplying with $\hbar$ (in meV\,ps) and
dividing by the Boltzmann constant $k_B$ (in meV/K).
In the fermionic case in Fig.~\ref{fig:example_ACE}(b), the Fermi energy is
specified by \texttt{Ferion\_E\_Fermi}.

The system-environment coupling operator $\hat{A}$ can be specified as 
matrix-valued expression using \texttt{Fermion\_SysOp} and 
\texttt{Boson\_SysOp}, respectively. 
The mode Hamiltonian is then set to
\begin{align}
\label{eq:DefSysOp}
H_E^{(k)} =& \sum_k \hbar \omega_k b^\dagger_k b_k + 
\sum_k \hbar g_k ( \hat{A} b^\dagger_k + \hat{A}^\dagger b_k),
\end{align}
for the \texttt{Boson} generator and equivalent with boson operators replaced 
by fermion operators for the \texttt{Fermion} generator.
This contains several often-used models: The default value for 
\texttt{Fermion\_SysOp} is \texttt{|0><1|\_2}, which corresponds to the 
resonant-level model describing a particle number conserving hopping
processes. The default value for \texttt{Boson\_SysOp} is \texttt{|1><1|\_2},
the projection onto the excited state, 
which describes a spin-boson model for a two-level system. The Jaynes-Cummings
model is obtained by setting \texttt{Boson\_SysOp} to \texttt{|0><1|\_2}.
Note that the  \texttt{Boson} generator automatically subtracts
the polaron shift or reorganization energy 
[see discussion of Eq.~\eqref{eq:spinboson}]. If this is not desired, it can 
be switched off by the command \texttt{Boson\_subtract\_polaron\_shift false}.

There are also several ways to specify the coupling constant to each of the
bath modes. If the coupling to all modes is identical, one can provide the
value of $g$ via \texttt{Fermion\_g}. However, if the modes 
discretize a continuum, it is instructive to instead supply the
rate $\Gamma$ expected in the Markov limit by Fermi's Golden Rule. This
is set by \texttt{Fermion\_rate} in Fig.~\ref{fig:example_ACE}(b) and
determines the coupling constants by 
$g=\sqrt{\Gamma(\omega_\textrm{max}-\omega_\textrm{min})/(2\pi N_E)}$.
Note that for any finite bandwidth $\omega_\textrm{max}-\omega_\textrm{min}$
the exact dynamics deviates from the Markovian dynamics---here $e^{-t}$---which 
is also shown in the results in Fig.~\ref{fig:example_ACE}(e).

When the system-environment coupling varies with frequency, one can instead
supply a spectral density defined by 
$J(\omega)=\sum_k |g_k|^2 \delta(\omega-\omega_k)$. This is used in the
the example of a spin-boson model in Fig.~\ref{fig:example_ACE}(c) and (f).
In the configuration file in panel (c), the command 
\texttt{Boson\_J\_from\_file}
instructs the code to read the spectral density from the file
\texttt{Boson.J}, in which we stored (frequency $\omega_j$ in the first column;
value of $J(\omega_j)$ in the second column) an ohmic spectral density
\mbox{$J(\omega)=0.2\, \omega \exp({-\omega/(3\textrm{ ps}^{-1})})$}.
The coupling to this bosonic environment results in damping of Rabi rotations as shown 
in Fig.~\ref{fig:example_ACE}(f).

\subsection{Selection of methods}
Whenever single modes or a corresponding generator is provided 
(the \texttt{...\_N\_modes} parameter set to a positive value),
the default behavior of the code is to calculate the corresponding 
PT-MPO using the ACE algorithm of Ref.~\cite{ACE}. 
As discussed in the method section~\ref{sec:methods} the ACE code
also supports the tree-like contraction scheme of Ref.~\cite{combine_tree},
which is used when the command \texttt{use\_combine\_tree true} is found 
in the configuration file.

The methods utilizing the Gaussian property of the spin-boson model
can be switched on by stating \texttt{use\_Gaussian true}. They then
process the same parameters as the \texttt{Boson} mode generator, such as the
spectral density, temperature, minimum and maximum frequency defining the frequency range,
and whether or not to subtract the polaron shift.
The default Gaussian method is the one by J{\o}rgensen and Pollock in Ref.~\cite{JP}.
The divide-and-conquer scheme of Ref.~\cite{DnC} is switched on by 
\texttt{use\_Gaussian\_divide\_and\_conquer true}.
Periodic PT-MPOs, also derived in Ref.~\cite{DnC}, are used when one sets
\texttt{use\_Gaussian\_periodic true}.

The Gaussian methods allow for memory truncation, i.e.~neglecting the bath correlation function beyond 
$n_\textrm{mem}=t_\textrm{mem}/\Delta t$ time steps. The memory time 
can be set in the configuration file by the parameter \texttt{t\_mem}.
Note that one should set the memory cut-off $n_\textrm{mem}$ to a power of two for the
divide-and-conquer as well as for the periodic PT-MPO method. 
Generally, the computation time does not monotonically decrease 
when the memory time is reduced. This is likely due to the fact that
a sudden jump to zero in the effective bath correlation function results in
spurious long-range temporal correlations that increase the inner bond
dimensions of the PT-MPOs. Heuristically, we suggest starting with a value of 
$t_\textrm{mem}$ which is about a factor 4 longer than the time scale 
on which the bath correlation function is found to drop to zero by
visual inspection, and then varying the memory time to find an 
optimum in computation time.

All the PT-MPO-based methods are implemented within the \texttt{ACE}
binary. Our framework further provides binaries \texttt{QUAPI} and 
\texttt{TEMPO}, which implement the path integral methods of 
Refs.~\cite{QUAPI1,QUAPI2,PI_cavityfeeding} and Ref.~\cite{TEMPO},
respectively. Both parse the same configuration file containing 
parameters of the \texttt{Boson} generator. Note that the memory
consumption of QUAPI scales exponentially with the number of 
memory time steps $n_\textrm{mem}$, so the corresponding parameter 
should not exceed $\sim 12$, even for small, e.g., two-level systems.

\subsection{\label{sec:fine_tuning}Convergence and fine tuning}
Several convergence parameters exist that control the accuracy as well as
the computation time of ACE and other PT-MPO techniques, such as the width 
of the time steps $\Delta t$ and MPO compression threshold $\epsilon$.
As described above, these are set via parameters \texttt{dt} and 
\texttt{threshold}.  
Details of the convergence depend on the chosen method and are discussed in the respective method papers~\cite{JP,ACE,DnC,combine_tree}. Here, we summarize the main insights and general trends. 

%

\begin{figure}
\includegraphics[width=\linewidth]{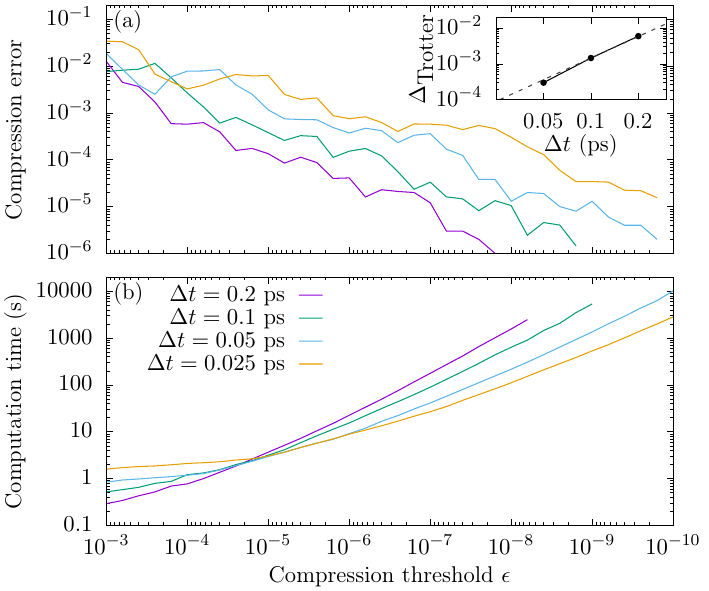}
\caption{\label{fig:convergence}Convergence for a spin-boson model as shown in Fig.~\ref{fig:example_ACE}(f). Panels (a) and (b) depict compression error and computation time, respectively, as a function of the nominal compression threshold $\epsilon$. Simulations are performed using the tree-like contraction of Ref.~\cite{combine_tree} and with exponentially increasing thresholds from $\epsilon/10$ to $\epsilon$ (\texttt{threshold\_range\_factor 10}). The inset in panel (a) shows the Trotter error obtained by comparing the maximal difference in the observable between the best converged (with respect to $\epsilon$) simulation for a given $\Delta t$ and reference calculation with time step $0.025$ ps. The dashed gray line the indicates a $\mathcal{O}(\Delta t^2)$ trend.} 
\end{figure}
An example for convergence is shown in Fig.~\ref{fig:convergence}, which depicts compression error, Trotter error, and computation times for the spin-boson model example of Fig.~ \ref{fig:example_ACE}(c) and (f). For these results, we modified the configuration file underlying Fig.~ \ref{fig:example_ACE}(c): First, we selected the tree-like contraction scheme by adding a line \texttt{use\_combine\_tree true}, second, we adjusted the number of modes to a power of two \texttt{Boson\_N\_modes 64} (as this is the optimal choice for the tree-like scheme) with fine tuning parameter  \texttt{threshold\_range\_factor 10} (see below). Finally, we proceeded to  vary the parameters \texttt{threshold} and \texttt{dt} and recorded the overall run times on a conventional laptop computer with AMD Ryzen 7 5825U processor.
Compression errors were obtained by comparing simulation results with larger threshold to results using the smallest threshold; dt was kept fixed in each such series of runs. Our definition of compression error is the maximal difference of occupations occurring at any time between $t=0$ and $t=t_e$.
Trotter errors were obtained by comparing with the simulations with smallest threshold in each series. 

First, it should be noted that the compression error for a given threshold $\epsilon$
strongly depends on $\Delta t$ and thus comparing simulations with
equal $\epsilon$ but different $\Delta t$ is not advised~\cite{ACE}. The maximal 
inner bond dimension of the respective final PT-MPOs tends to be a more stable
indicator of the absolute compression error when the time steps $\Delta t$ are
changed~\cite{ACE}. The inner bonds can be extracted from a process tensor
file using the binary 
\texttt{PTB\_analyze -read\_PT FILE.pt}.
Thus, to gauge the convergence with respect to both parameters, it is instructive
to perform simulations where, for several values of the time step $\Delta t$, a 
series of values for the threshold spanning several orders of magnitude are tested
and the impact on the observables is checked. Only after the compression error for fixed $\Delta t$ is
understood, the Trotter error due to the finite time step $\Delta t$ can be checked reliably. 

In the example in Fig.~\ref{fig:convergence}, we see that to achieve results converged to a Trotter error of about $10^{-3}$, a time step $\Delta t\lesssim 0.1$ ps is required. For $\Delta t=0.1$ ps, a threshold of the order $\epsilon \lesssim 10^{-5}$ yields a compression error comparable to the Trotter error. The corresponding calculations finish within a few seconds. Extrapolating the $\mathcal{O}(\Delta t^2)$ trend, we estimate a Trotter error of $\sim 10^{-4}$ for $\Delta t\sim 0.025$ ps, for which a similar compression error is achievable within a few minutes. 

Moreover, if the environment modes in the ACE algorithm arise from discretizing 
a continuum, the number of modes $N_E$ (e.g., \texttt{Boson\_N\_modes}) and the bandwidth 
$\omega_\textrm{max}-\omega_\textrm{min}$ (e.g., \texttt{Boson\_omega\_max} and 
\texttt{Boson\_omega\_min}) constitute additional convergence parameters.
For a given bandwidth, the optimal value of $N_E$ depends on the total propagation time
$t_e-t_a$ via energy-time uncertainty. We recommend the choice 
$N_E=0.4 (\omega_\textrm{max}-\omega_\textrm{min}) (t_e-t_a)$, where $0.4$ is a heuristic
factor~\cite{ACE}. 
Note also that increasing the number of modes for a fixed bandwidth results in
weaker coupling $g_k$ per mode. This again affects how the compression error on 
observables scales with the threshold $\epsilon$, and thus simulations with 
equal thresholds but different number of modes per bandwidth are not directly 
comparable.

For methods relying on the preselection approach for the PT-MPO combination,
it is important to keep in mind that the corresponding compression can be significantly suboptimal,
resulting in larger inner bond dimensions compared to other methods and also 
in more severe error accumulation. Especially in cases where very small time steps are used,
simulations have been found to not converge with decreasing threshold~\cite{DnC,combine_tree}.
However, this can be mitigated by fine-tuning the compression. 

To this end, several strategies have been explored:
First, the divide-and-conquer and the periodic PT-MPO methods of Ref.~\cite{DnC}
often profit from using a smaller threshold for singular value selection 
and backward sweep compared to the forward sweep. This is because the selection
and backward sweep provides a less controlled truncation than the forward sweep,
and the forward sweeps with coarser threshold partially remove spurious singular
values introduced by the former. While the parameter \texttt{threshold} is 
used as a base value for the threshold, with parameters 
\texttt{forward\_threshold\_ratio} and \texttt{backward\_threshold\_ratio} one
can set different thresholds for the two directions relative to the base value.
Moreover, the parameter \texttt{select\_threshold\_ratio} specifically 
changes the threshold used in the preselection step. For example, in 
Ref.~\cite{DnC}, we found \texttt{backward\_threshold\_ratio 0.2} 
to result in reduced computation times for PT-MPOs describing the effects
of phonons on semiconductor quantum dots.

For the tree-like ACE contraction scheme~\cite{combine_tree}, we found it
beneficial to employ a dynamically increasing threshold, where we keep the same
thresholds for forward and backward sweeps within each pair of sweeps after a PT-MPO combination step
but we gradually increase the threshold as the PT-MPO grows. 
Specifically, we start with a small threshold $\epsilon_\textrm{min}=\epsilon/r$ for the first PT-MPO combination
and exponentially increase (linear interpolation of $\log \epsilon$) the threshold such that 
the compression after the final combination step occurs with threshold 
$\epsilon_\textrm{max}=\epsilon$. The factor $r$ is specified by
\texttt{threshold\_range\_factor}. A value of $r=10$ to $r=100$ 
is often useful~\cite{combine_tree}. This fine-tuning strategy turns out to be useful also
for Gaussian PT-MPO methods (see example in Sec.~\ref{sec:G2}).

\subsection{Trotter Errors}
The decomposition of the total propagator into system and envrionment
parts in Eq.~\eqref{eq:explicit_evolution} has been derived using the
asymmetric (first-order) Trotter decomposition 
$e^{\mathcal{L}\Delta t} =e^{\mathcal{L}_E\Delta t} e^{\mathcal{L}_S\Delta t}+\mathcal{O}(\Delta t^2)$
in Eq.~\eqref{eq:asymTrotter}. The Trotter error can be reduced by
using instead the symmetric (second order) Trotter decomposition
$e^{\mathcal{L}\Delta t} =e^{\mathcal{L}_S\Delta t/2} e^{\mathcal{L}_E\Delta t} e^{\mathcal{L}_S\Delta t/2}+\mathcal{O}(\Delta t^3)$.
The use of the symmetric Trotter decomposition is now in fact the default behavior
setting in our code. 

However, for simulations with more than one PT-MPO with mutually non-communiting  
interaction Hamiltonians, we suggest an alternative decomposition, 
where the order of applying the PT-MPO matrices (with respect to outer index 
multiplication) alternates\footnote{Trotter decompsitions with alternating order were also proposed, e.g., for the modular path integral method in Ref.~\cite{ModularPI}}. Here, the understanding of PT-MPO matrices as compressed
environment propagators is useful~\cite{inner_bonds}. For example, if there are two
environment propagators $e^{\mathcal{L}_E^{(1)}\Delta t}$ and
$e^{\mathcal{L}_E^{(2)}\Delta t}$ in addition to the system propagator
$e^{\mathcal{L}_S\Delta t}$, the propagation over two time steps 
is given by 
\begin{align}
e^{\mathcal{L}_S\Delta t}e^{\mathcal{L}_E^{(1)}\Delta t}e^{\mathcal{L}_E^{(2)}\Delta t}
e^{\mathcal{L}_E^{(2)}\Delta t}e^{\mathcal{L}_E^{(1)}\Delta t}e^{\mathcal{L}_S\Delta t},
\end{align}
which describes a symmetric Trotter decomposition of the joint environment propagator
over two time steps 
\begin{align}
e^{(\mathcal{L}_E^{(1)}+\mathcal{L}_E^{(2)})(2\Delta t)}=
e^{\mathcal{L}_E^{(1)}\Delta t}e^{\mathcal{L}_E^{(2)}(2\Delta t)}e^{\mathcal{L}_E^{(1)}\Delta t}
+\mathcal{O}(\Delta t^3)
\end{align}
followed by a symmetric Trotter decomposition between system and the joint environment
\begin{align}
&e^{(\mathcal{L}_S+\mathcal{L}_E^{(1)}+\mathcal{L}_E^{(2)})(2\Delta t)}
\nonumber\\&=   
e^{\mathcal{L}_S\Delta t}e^{(\mathcal{L}_E^{(1)}+\mathcal{L}_E^{(2)})(2\Delta t)}e^{\mathcal{L}_S\Delta t}
+\mathcal{O}(\Delta t^3).
\end{align}

This feature is turned on in the code on by setting the parameter \texttt{propagate\_alternate true},
which overrides the use of the symmetric Trotter decomposition. Alternating the propagation order
can lead to zigzagging behavior in the output, as observables at odd time steps have a larger Trotter error
order than at even time steps. One should then keep only the values at even time steps (counting from zero). 
The difference between odd and even time steps can be used as an indicator for the Trotter error.

\subsection{\label{sec:QDcavity}Example: Composite system of QD and microcavity}
\begin{figure*}
\includegraphics[width=\linewidth]{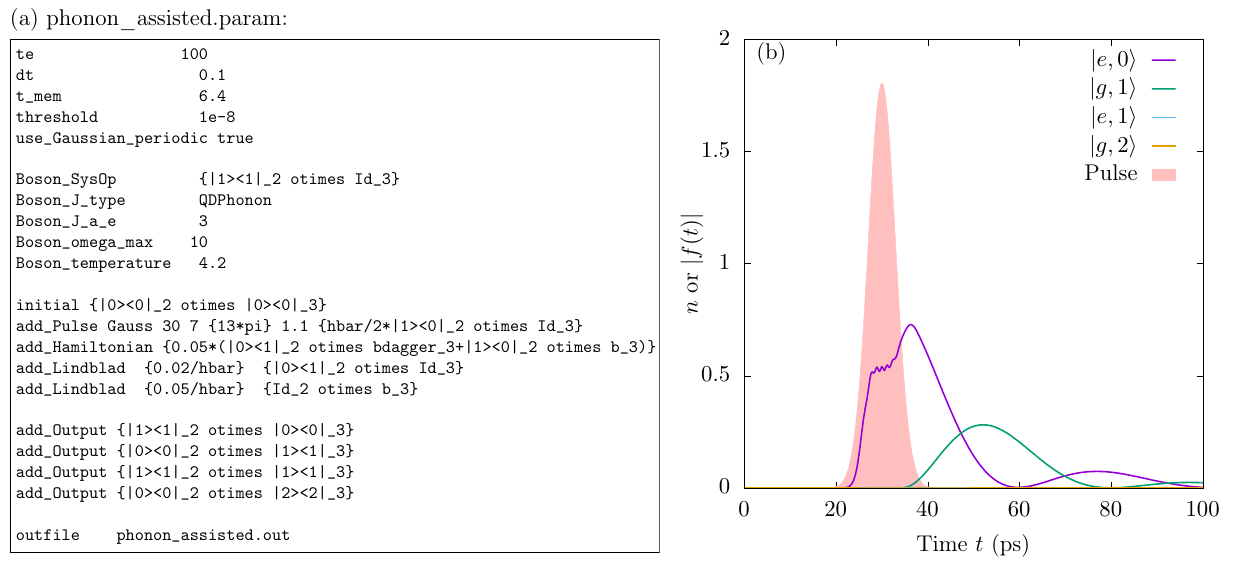}
\caption{\label{fig:phonon_assisted}(a) Configuration file and (b) simulation
results for phonon-assisted single-photon generation using a QD
strongly coupled to an optical microcavity.}
\end{figure*}
We now provide further examples to demonstrate some of the features mentioned 
previously.  
First, we consider a bipartite open quantum system, namely a
semiconductor QD strongly coupled to 
longitudinal acoustic phonons as well as strongly coupled to an 
optical single-mode microcavity. The QD and the cavity are treated as part of the system.
The phonon environment only couples to the QD part of the system.
For semiconductor QDs, the coupling between electronic excitations and phonons
can be derived from microscopic considerations~\cite{Krummheuer}. 
In particular, assuming infinite-potential confinement along the growth direction 
and parabolic confinement in the in-plane directions, the electron-phonon
coupling gives rise to a spin-boson model with spectral density
\begin{align}
\label{eq:krummheuer_SD}
J(\omega)=&\frac{\omega^3}{4\pi^2\rho\hbar c_s^5}
\bigg(D_e e^{-\omega^2a_e^2/(4c_s^2)} - D_he^{-\omega^2a_h^2/(4c_s^2)}\bigg)^2,
\end{align}
where, for a QD in a GaAs matrix, the mass density is $\rho=5370$ kg/m$^3$, 
the speed of sound is $c_s=5110$ m/s, and the electron and hole deformation 
potential constants are $D_e=7.0$ eV and $D_h=-3.5$ eV, respectively~\cite{Krummheuer}.
The lengths $a_e$ and $a_h$ are the electron and hole radii, respectively.

Because of the importance of QDs for quantum technology, we have implemented as a
convenience option the specification of the spectral density in Eq.~\eqref{eq:krummheuer_SD}.
As show in the configuration file in Fig.~\ref{fig:phonon_assisted}(a), 
this spectral density is chosen by setting \texttt{Boson\_J\_type QDPhonon}. 
The electron and hole radii can be set by \texttt{Boson\_J\_a\_e} and 
\texttt{Boson\_J\_a\_h}, respectively. If not set explicitly, $a_e$ has
the value 4 nm and $a_h=a_e/1.15$.
Moreover in Fig.~\ref{fig:phonon_assisted}(a) we set 
\texttt{use\_Gaussian\_periodic true} to generate a periodic PT-MPO. This
requires a memory cutoff $n_\textrm{mem}$ which is a power of 2 times the time step $\Delta t=0.1$ ps,
for which we set the parameter \texttt{t\_mem} to 6.4 ps.

The concrete situation modeled in Fig.~\ref{fig:phonon_assisted} is phonon-assisted
state preparation for a QD strongly coupled to a microcavity, reproducing the 
example in Ref.~\cite{PI_singlephoton}. There, the coupled system is driven by
a blue-detuned (1.1~meV with respect to the two-level transition energy) Gaussian laser pulse.
The cavity mode is on resonance with the two-level system. 
The QD-cavity coupling has a strength of $\hbar g=0.05$ meV, the cavity is lossy 
with loss rate \mbox{$\kappa=0.05 \textrm{ meV}/\hbar$}, and two-level excitations decay
radiatively with loss rate \mbox{$\gamma=0.02 \textrm{ meV}/\hbar$}.

Note that the system initial state, Hamiltonians (pulses), Lindblandians, observables,
and the system-environment coupling operator \texttt{Boson\_SysOp} all 
act on the 6-dimensional composite system Hilbert space containing both QD and the 
truncated cavity mode. However, because the system-environment coupling operator
is highly degenerate---which is automatically identified by the code---the PT-MPO
calculation is only as difficult as that for an isolated two-level system (see Sec.~\ref{sec:outer_reduction} Outer bond reduction).

The dynamics depicted in Fig.~\ref{fig:phonon_assisted}(b) can be understood 
in terms of adiabatic undressing~\cite{adiabatic_undressing}: The strong 
pulse leads to laser-dressing of states, such that (i) phonon-assisted
transitions between dressed states are possible and lead to fast
thermalization towards the lower dressed state, and (ii) the lower
dressed states adiabatically evolves into the excited state as the 
pulse vanishes. A corresponding jump in the occupations of state 
$|e,0\rangle$ (excited states with zero photons in the cavity) is observed at the end of the pulse.
The excitation is then transferred to the cavity via the QD-cavity coupling and then out-coupled from
the cavity via cavity losses.
Moreover, during the pulse, the dressing detunes the QD from the cavity 
frequency, which suppresses emission during the pulse and thereby reduces
reexcitation. Because the laser is also off-resonant from the cavity,
the phonon-assisted scheme combines several important features of
single-photon generation: high single-photon purity, seperability of
emitted photons from stray laser photons, and relatively high 
brightness. Because of these advantages, the phonon-assisted scheme
is used in practical implementations of single-photon 
sources~\cite{Thomas_singlephoton}.

\subsection{\label{sec:G2}Example: Photon coincidences from superradiant QDs}
\begin{figure*}
\includegraphics[width=\linewidth]{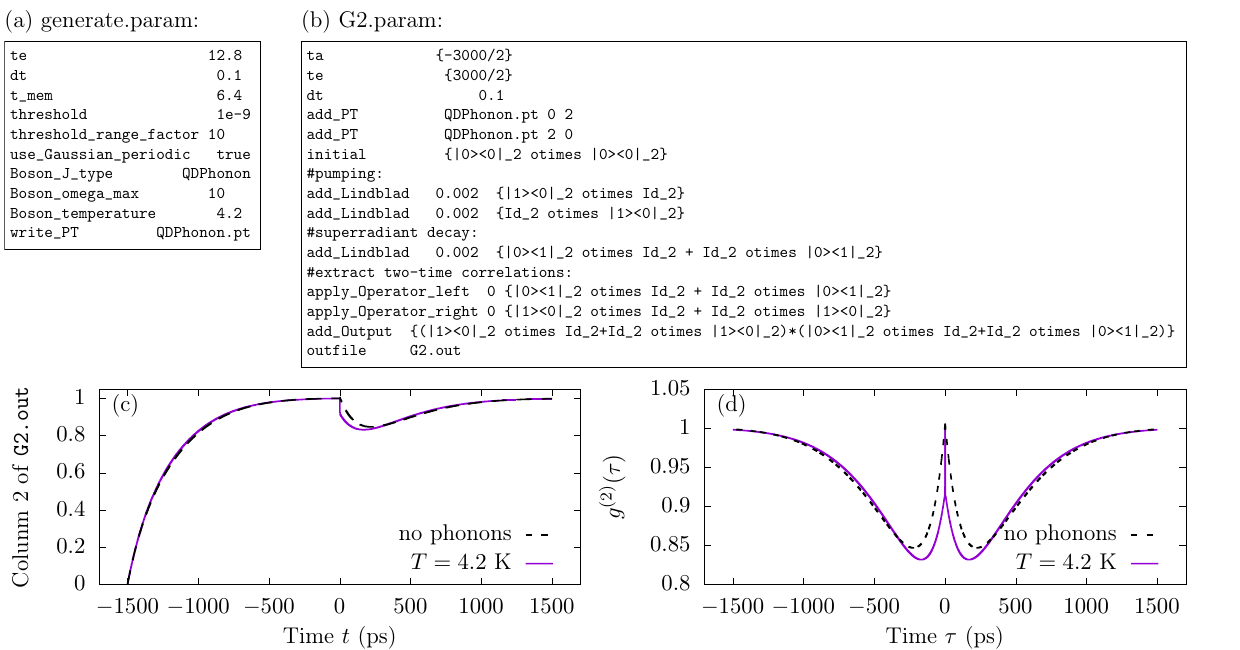}
\caption{\label{fig:G2}Configuration files for (a) PT-MPO
generation and (b) simulation of photon coincidences from
two incoherently driven superradiant quantum strongly coupled
to local phonon baths. (c) Generated output with and 
without accounting for phonons. (d) Photon coincidences.
}
\end{figure*}
A further example illustrates the use of multiple PT-MPOs at the same time.
Motivated by recent experiments, which demonstrated cooperative emission
from indistinguishable QDs~\cite{Waks_superradiance,Gammon_superradiance,
CoopSciAdv, Lodahl_superradiance},
we consider two QDs each coupled to a local non-Markovian phonon bath 
and both QDs coherently coupled to the electromagnetic environment. 

Assuming a flat spectral density of the eletromagnetic environment,
the latter can be described by a Lindblad term for the joint radiative
decay~\cite{Jacobs_Lindblad,WiercinskiPolaronDressing}. If the QDs 
have identical energies, the radiative decay is enhanced with respect
to the emission from individual or distinguishable QDs, which is described
by Lindbald terms involving the symmetric linear combination of dipole 
operators~\cite{CygorekCoop} $2 \kappa \mathcal{D}_{\sigma^-_S}[\bar{\rho}]$,
where
\begin{align}
\sigma^{\pm}_S=\frac 1{\sqrt{2}}( \sigma^\pm_1 +\sigma^\pm_2)
\end{align}
and $\kappa$ is the radiative decay of a single QD. We further assume
the QDs to be independently pumped with a pump rate $\gamma_p$, which is
described by Lindbladians $\gamma_p\mathcal{D}_{\sigma_1^+}[\bar{\rho}] 
+\gamma_p\mathcal{D}_{\sigma_2^+}[\bar{\rho}]$. Here, we set
$\kappa=\gamma_p=1/(0.5\textrm{ ns})$.

To reveal collective effects in few-emitter systems, one often measures
photon coincidences~\cite{Waks_superradiance,Gammon_superradiance,CoopSciAdv}
\begin{align}
g^{(2)}(t,\tau)=& \frac{G^{(2)}(t,\tau)}{I(t)I(t+\tau)},
\end{align}
where $I(t)=\langle \sigma^+_S(t) \sigma^-_S(t) \rangle$ is the intensity
and 
\begin{align}
&G^{(2)}(t,\tau)=\langle \sigma^+_S(t)\sigma^+_S(t+\tau)\sigma^-_S(t+\tau)\sigma^-_S(t)\rangle
\nonumber\\&=\textrm{Tr}\big[ \big(\sigma^+_S(t+\tau)\sigma^-_S (t+\tau)\big)\big( \sigma^-_S(t) \rho(0) \sigma^+_S(t)\big)  \big]
\label{eq:G2}
\end{align}
are the unnormalized coincidences. From the last line of Eq.~\eqref{eq:G2}, it is 
clear that $G^{(2)}$ can be obtained by propagating the open quantum system
over a time $t$, then applying operator $\sigma^-_S$ from the left and 
$\sigma^+_S$ from the right, propagating further over time $\tau$, and finally evaluating
the observable $\sigma^+_S\sigma^-_S$.
This is precisely how we evaluate $G^{(2)}$ using the ACE code with the configuration
file in Fig.~\ref{fig:G2}(b), where we chose $t=0$.

Note that of the output, which is shown in Fig.~\ref{fig:G2}(c), only the data
points at strictly positive times $\tau>0$ are related to $G^{(2)}$ while 
data points at previous time steps describe the evolution of the 
intensity $I(t)$. Therefore, in Fig.~\ref{fig:G2}(d), we mirror the results to 
also plot coincidences for negative delay times $\tau$ and zoom into the relevant range.

To account for phonon effects, we add two PT-MPOs to the simulation. This is done
using the \texttt{add\_PT} command, whose first argument is the name of the
corresponding PT-MPO file, while the second and third arguments are optional parameters
that denote whether the outer bonds of the PT-MPOs shall be temporarily 
(no change to file occurs) extended to support a larger composite system Hilbert
space. Concretely, outer bonds calculated for a Hilbert space $\mathcal{H}^{(0)}_S$
are extended to support a composite space
$\mathcal{H}_\textrm{left}\otimes\mathcal{H}^{(0)}_S \otimes\mathcal{H}_\textrm{right}$, 
where the second argument of the \texttt{add\_PT} command denotes
the dimension of $\mathcal{H}_\textrm{left}$ and the third argument is the
dimension of $\mathcal{H}_\textrm{right}$.
Hence, the first line in Fig.~\ref{fig:G2}(b) containing \texttt{add\_PT} 
indicates the PT-MPO that acts on the first QD 
and the next line describes the PT-MPO acting of the second QD.

Here, we use the same PT-MPO for both QDs, which is precalculated using the 
configuration file in Fig.~\ref{fig:G2}(a). Again, we calculate a periodic PT-MPO and
employ fine-tuning using the threshold range factor $r=10$ to slightly reduce
the bond dimension. The command \texttt{write\_PT} instructs the ACE code to
write the calculated PT-MPO to the corresponding file. 
Generally, more than one PT-MPO file may be created, whose names start with the 
provided name. This is done to facilitate buffering, i.e. reading and writing 
from and to files. For example, a PT-MPO may be split up into several files each
containing $B$ blocks of the PT-MPO using the command \texttt{buffer\_blocksize}
followed by the number $B$. This is useful when the full PT-MPO does not fit
into working memory. Instead of \texttt{add\_PT}, one can also
load PT-MPOs using \texttt{initial\_PT}. The difference is that the latter can only
use at most one PT-MPO and potentially modifies it. For example, using  
\texttt{initial\_PT}, combined with \texttt{add\_single\_mode}, and
\texttt{write\_PT} modifies an existing PT-MPO to include the effects of another
single environment mode.

The explanation of the physics of the results in Fig.~\ref{fig:G2}(d) follows along
the lines of the analysis in Refs.~\cite{CygorekCoop, CoopWiercinski, WiercinskiPolaronDressing}.
First, photon coincidences without phonons can be derived analytically~\cite{CygorekCoop}. 
The exact dynamics and the value of 
\mbox{$g^{(2)}(\tau=0)$} depend on the details, such as the ratio between 
pump strength and radiative decay. In any case, one observes a peak with a value
of \mbox{$g^{(2)}(\tau=0)$} significantly larger than 0.5, which is the limit for
photon coincidences from distinguishable, uncorrelated emitters. The excess 
$g^{(2)}$ is directly related to inter-emitter coherences.
Second, when a two-level system subject to a spin-boson interaction 
is not driven (no Hamiltonian term), one obtains an independent-boson model, 
which can be solved analytically~\cite{BreuerPetruccione}.
For an independent-boson model with super-Ohmic spectral density, 
coherences initially drop but then remain constant. 
This behavior of the two-level system also translates to inter-emitter
coherences of QDs coupled to two phonon baths~\cite{CoopWiercinski}, 
where phonons are found to also result in an initial fast drop in $g^{(2)}$ but 
after a few ps the dynamics with phonons decays and restores in parallel with
the phonon-free case. 

This example demonstrates how the ACE code can solve 
a multi-partite, non-Markovian multi-environment, and multi-scale problem 
in a numerically complete way with only few lines in the configuration files
and no further programming required.

\subsection{\label{sec:structured}Example: Strongly structured environment}
\begin{figure}
\includegraphics[width=\linewidth]{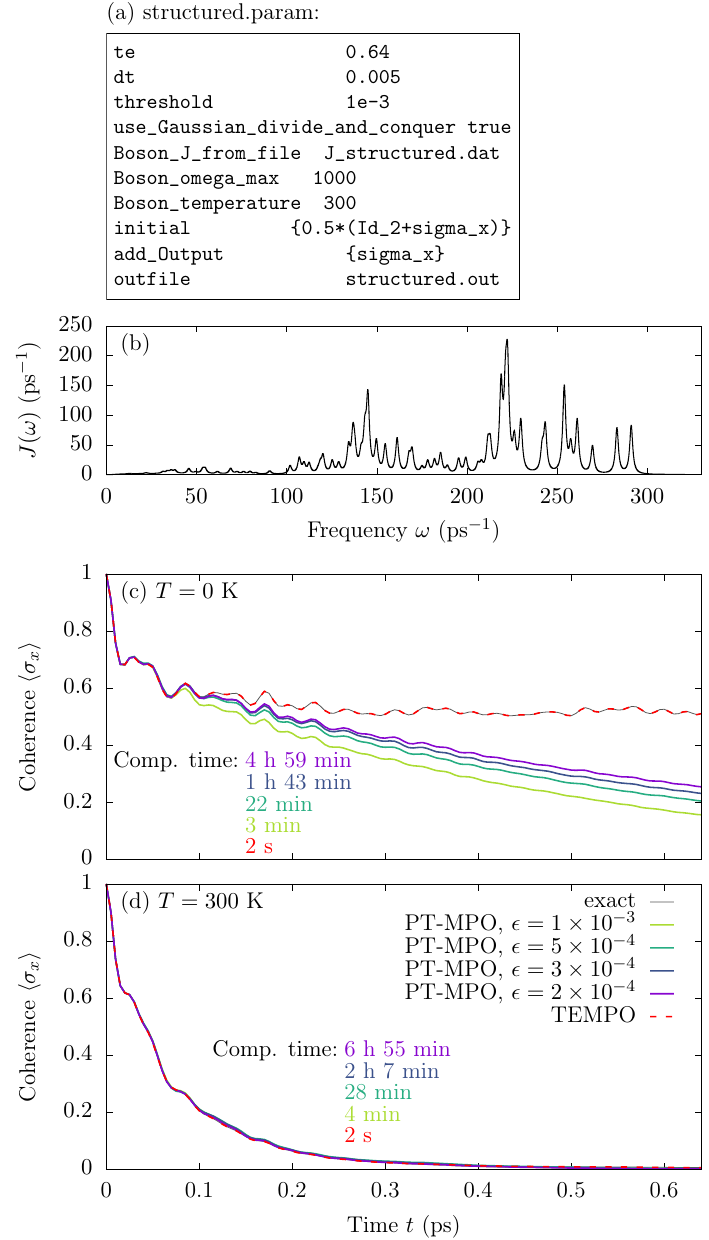}
\caption{\label{fig:structured}Simulation of the free decay of initial coherences in a two-level system that is coupled to an environment featuring the strongly structured spectral density of the FMO complex~\cite{Ratsep}. Panel (a) shows the configuration file and panel (b) the spectral density, while (c) and (d) depict the dynamics of the coherences for temperature $T=0$ and $T=300$ K, respectively. PT-MPO simulations with different thresholds are compared with the exact dynamics obtained from polaron transformation as well as with TEMPO simulations.}
\end{figure}
Finally, we consider an example that is particularly challenging for PT-MPO-based methods. Because the inner bonds of PT-MPOs account for the dynamics within the most relevant environment degrees of freedom~\cite{inner_bonds}, the bond dimension becomes large when the system is coupled to a large number of modes, which appear as a series of narrow peaks in a strongly structured spectral density. Such a situation arises, e.g., in the Fenna-Matthews-Olson (FMO) photosynthetic complex, where a spectral density consisting of 62 modes was estimated from experiments in Ref.~\cite{Ratsep}. Here, we use the same spectral density as Lorenzoni \textit{et al.}~in Ref.~\cite{Lorenzoni_PRL}, where the 62 sharp $\delta$-like resonances are described by slightly broadened Lorentzians, and an Adolphs-Renger contribution is added to provide a low-energy background [see Fig.~\ref{fig:structured}(b)].

We couple a spin-boson environment with this spectral density to a two-level system and observe the free decay of coherences. Specifically, the system is initialized in an equal superposition of ground and excited state, we adopt the trivial system Hamiltonian $H_S=0$, and we then extract the expectation value of the observable $\sigma_x$.
The narrow peaks in the spectral density lead to long memory times, while at the same time the high frequencies of some of the modes necessitates resolving fast oscillations with small time steps. The most advantageous PT-MPO method is then the divide-and-conquer scheme of Ref.~\cite{DnC} without memory truncation. The corresponding configuration file is shown in Fig.~\ref{fig:structured}(a). 

The resulting dynamics of the free coherence decay is depicted in Fig.~\ref{fig:structured}(c) and (d) for temperatures $T=0$ and $T=300$ K, respectively, and for different values of the threshold $\epsilon$ (requiring only the modification of the parameters \texttt{Boson\_temperature} and \texttt{threshold} in the configuration file). The results are compared with the exact expression derived from polaron transformation~\cite{Review_Nazir} (which is available courtesy of our choice $H_S=0$)
\begin{align}
\label{eq:structured_exact}
&\langle \sigma_x(t)\rangle = e^{\int\limits_0^\infty d\omega\;\frac{J(\omega)}{\omega^2}\big[(\cos \omega t -1) \coth(\beta\omega/2)- i \sin \omega t\big]}.
\end{align}
At low temperature ($T=0$), the exact time evolution of the coherence shows an initial drop within the first $\sim 100$ fs, after which it fluctuates around a nearly constant value. The PT-MPO simulations match well with this behavior during the initial 100 fs but convergence out to larger times becomes very slow and challenging. Even for a relatively large threshold of $\epsilon=3\times 10^{-4}$ we find a PT-MPO with bond dimensions of $\sim 1000$, which entails a computation time of a few hours. Note that such a large bond dimension is not due to the particular method (nor its specific implementation) but must occur as an intrinsic feature of a PT-MPO: As PT-MPOs are entirely independent of the system Hamiltonian, they must be able to describe the environment response to all possible system drivings. Driving the system at a frequency that is resonant with any environmental mode leads to significant excitation of---and back-action from---that particular mode. Correspondingly, the inner bonds of the PT-MPO must accommodate the description of the large number of configurations of environment excitations that is reachable with arbitrary system driving. 

By contrast, TEMPO~\cite{TEMPO} is a tensor network method that efficiently encodes temporal correlations within the state of the system for a given system Hamiltonian. 
This method is implemented in our framework with the binary \texttt{TEMPO}, which processes the same configuration files as \texttt{ACE}. The TEMPO results obtained with the configuration file shown in Fig.~\ref{fig:structured}(a) match the exact expression in Eq.~\eqref{eq:structured_exact} perfectly. Due to the trivial system Hamiltonian $H_S=0$, temporal correlations in the states are simple, and the computation time is negligible. 
It should, however, be noted that for more general $H_S$ and environments with less structure, PT-MPO-based methods are typically significantly faster than TEMPO, see, e.g., examples in Ref.~\cite{JP} and the supplementary material of Ref.~\cite{ACE}. Which method is more advantageous thus strongly depends on the details of the physical system at hand.

Finally, as shown in Fig.~\ref{fig:structured}(d), for the free decay of coherences at physiological temperatures $T=300$ K, we find immediate agreement between the different methods. The computation times reflect the fact that the PT-MPOs have even larger bond dimensions compared with the respective simulation at zero temperature. However, because the coherences are strongly damped and decay to zero with a decay time of $\sim 60$ fs (obtained from an exponential fit), deviations to the exact result are also strongly suppressed. It should be noted that the overall error in general depends not only on the nominal threshold but also on the system Hamiltonian and on the chosen observable. For example, we observe that deviations from trace preservation eventually accumulate to several percent (not shown) even for the smallest threshold in PT-MPO simulations at $T=300$ K. Thus, TEMPO simulations also provides a useful reference in the high-temperature regime. 

\section{Summary}
We have described the ACE code~\cite{ACEcode}, which is a versatile solver for non-Markovian
open quantum systems based on PT-MPOs. The concrete physical problem is specified
in configuration files. The corresponding commands and parameters are discussed
on a series of examples from simple closed systems to multi-partite 
multi-environment problems. 

The ACE code implements several methods to calculate PT-MPOs and allows PT-MPOs
to be stored in files, manipulated, combined with other PT-MPOs, and read again 
to efficiently scan simulation results for different time-dependent system
Hamiltonians. 
When the environment is of the form of a Gaussian spin-boson model, one
can use the algorithm by J{\o}rgensen and Pollock~\cite{JP}, the divide-and-conquer
algorithm of Ref.~\cite{DnC}, or periodic PT-MPOs. If the environment is more 
generally composed of independent modes, the ACE algorithm of Ref.~\cite{ACE} 
as well as an enhanced version with a tree-like combination scheme~\cite{combine_tree}
is available. Moreover, it is possible to use 
several PT-MPOs to describe multiple distinct environment influences.
This makes the code extremely versatile, and it demonstrates the core idea for
the development of a universal numerically exact solver for networks of 
non-Markovian open quantum systems based on PT-MPOs.

A current limitation of the code is the size of the open quantum system that can be tackled. For example, the PT-MPO algorithm by J{\o}rgensen and Pollock~\cite{JP} nominally scales as $\mathcal{O}(D^8)$ with the system dimension $D$\footnote{Exact SVD decomposition algorithms of a $N\times M$ matrix with $N>M$ generally scales as $\mathcal{O}(NM^2)$. For a single row of the tensor network in the J{\o}rgensen-Pollock approach~\cite{JP}, $M$ corresponds to the inner bonds dimension $D^2$, while $N=M D^2$ additionally contain the outer bonds. This yields $\mathcal{O}(D^8)$ overall}. For non-Gaussian environments with general off-diagonal coupling to modes of Hilbert space dimension $D_M$, the nominal scaling of single SVDs in ACE is $\mathcal{O}(D^4 D_M^6)$.
However, whenever the system-environment coupling operator has degeneracies, such as for composite systems, the outer bond dimension is drastically reduced as described in Sec.~\ref{sec:outer_reduction}. For Gaussian methods, the automatic detection of degeneracies also reduces in inner bond dimension of the initial tensor network. This way, it was possible to use our ACE code for simulations of open quantum systems with over 30 levels, e.g., 4 levels of a QD coupled to two bosonic modes truncated each at Hilbert space dimension 3 in Ref.~\cite{SUPER_entanglement} and superradiant emission from 5 closely spaced quantum emitters in Ref.~\cite{DnC}.
For situations where no such degeneracies can be exploited, an approach along the lines of Ref.~\cite{QASPEN} may enable the simulation of larger system by systematically approximating the coupling Hamiltonian by a lower-rank operator using Chebyshev interpolation.

It is worth stressing once more that 
a key feature distinguishing PT-MPOs from other approaches is the former's independence of the system propagator.
On the one hand, this enables fast scans of system parameters with a PT-MPO that has to be calculated only once, the evaluation of arbitrary multi-time correlation functions, and the numerically exact simulation for systems with multiple non-Markovian environments. On the other hand, this feature prohibits PT-MPO methods from utilizing assumptions about the system dynamics, which would limit the complexity of the allowed environment response. This is manifested in the fact that the inner bond dimensions $\chi$ of PT-MPOs play a crucial role resulting in an additional $\mathcal{O}(\chi^3)$ scaling, e.g., for system propagation, where the value of $\chi$ strongly depends on the concrete environment and is difficult to estimate a priori. Thus, for large spin-boson systems coupled to strongly structured environments (and consequently large bond dimensions) and with a fixed, time-independent system Hamiltonians, methods based on the construction of effective propagators operating only on the reduced system Liouville space like SMatPI~\cite{SMatPI} likely outperform PT-MPOs. For time-independent Hamiltonians, effective propagators for long-time dynamics can also be extrapolated from short-time propagators (over about $n_\textrm{mem}$ time steps) from any non-Markovian open quantum systems approach by post-processing using the Transfer Tensor Method (TTM)~\cite{TransferTensor}.
To offer access to the improved performance that can become available once a system Hamiltonian has been fixed, our code also provides implementations of QUAPI and TEMPO as alternatives to PT-MPOs.

Finally, note that the propagation of multi-partite system with several PT-MPOs 
by the analog of Eq.~\eqref{eq:iter} involves the multiplication of matrices,
whose dimensions are the products of the individual system Liouville space 
dimensions and the inner bonds of the PT-MPOs. Future work will be directed 
towards tackling the exponential scaling in simulations of quantum networks with respect to the number of constituent parts. 
The combination of PT-MPOs with many-body techniques like TEBD in Ref.~\cite{Fux_spinchain} exemplifies promising efforts in this direction.

\acknowledgements
We are grateful for fruitful discussions with Jonathan Keeling, Brendon W. Lovett, Julian Wiercinski, and Thomas Bracht.
M.C. is supported by the Return Program of the State of North Rhine-Westphalia. M.C.~and
E.M.G.~acknowledge funding from EPSRC grant no. EP/T01377X/1. 

\appendix
\section{\label{app:commands}Summary of commands and arguments}
\newcommand{\tabsubtitle}[1]{\hline
\multicolumn{3}{|l|}{\textbf{#1}}\\
\hline}
\begin{table*}
\renewcommand{\arraystretch}{1.25}
\begin{tabularx}{\linewidth}{|l |l |X|}
\hline
Command & Arguments & Comments \\
\tabsubtitle{Basic controls:}
\texttt{dt} & float & Time step $\Delta t$ (unit: ps, default value: 0.01)\\
\texttt{ta} & float & Initial time of simulation $t_a$ (unit: ps, default value: 0) \\
\texttt{te} & float & Final time of simulation $t_e$ (unit: ps, default value: 10) \\
\texttt{outfile} & string & Name of output file to be created\\
\texttt{use\_symmetric\_Trotter} & bool & Switches on second order (as opposed to first-order) Trotter splittings between system propagators and PT-MPO matrices (default value: true). \\
\texttt{propagate\_alternate} & bool & Switches on alternating order for system and envrionment propagators for multi-environment simulations (overrides \texttt{use\_symmetric\_Trotter}; default value: false).\\
\texttt{set\_precision} & int & Changes number of significant digits of floating point numbers written to \texttt{outfile} \\
\tabsubtitle{System parameters:} 
\texttt{initial} & matrix & Initial system density matrix.\\
\texttt{add\_Hamiltonian} & matrix & Adds [argument] to the (time-constant) system Hamiltonian. \\
\texttt{add\_Pulse} & string  [...] & Adds time-dependent part of system Hamiltonian.  The first arguments determines the type of pulse, e.g., \texttt{file} for reading pulses from a file or \texttt{Gauss} for using a predefined Gaussian pulse. The remaining parameters depend on this choice [see example in Fig.~\ref{fig:Markov}(b)].
\\
\texttt{add\_Lindblad} & float  matrix & Adds a Lindblad term to the free system propagator. The first argument is the rate in ps$^{-1}$, the second argument is the collapse operator. \\
\texttt{apply\_Operator\_left} & float matrix & Multiplies the system density matrix at time [first argument] with an operator [second argument] from the left, e.g., to extract multitime correlation functions.\\ 
\texttt{apply\_Operator\_right} & float matrix &Multiplies the system density matrix at time [first argument] with an operator [second argument] from the right.\\
\texttt{add\_Output} & matrix & Specifies an observable [argument] to be extracted from the reduced system density matrix. Every occurrence of \texttt{add\_Output} adds two columns to the \texttt{outfile}, corresponding to real and imaginary parts of the observable.\\
\tabsubtitle{Handling of PT-MPOs and compression:} 
\texttt{threshold} & float & Base MPO compression threshold; relative to largest SVD (default value: 0=no truncation) \\
\texttt{t\_mem} & float & Memory time used for memory truncation $t_\textrm{mem}$ (ps) \\
\texttt{n\_mem} & float & Memory cut-off (number of steps) used for memory truncation. Overrides \texttt{t\_mem}.  \\
\texttt{threshold\_range\_factor} & float & The threshold is multiplied by a factor, which is exponentially interpolated from $1/r$ for the first MPO compression sweep to $1$ for the final compression sweep.\\
\texttt{forward\_threshold\_ratio} & float & When sweeping in forward direction (from $t=t_a$ to $t=t_e$), the compression threshold is multiplied by this value (default value: 1).\\
\texttt{backward\_threshold\_ratio} & float & When sweeping in backward direction (from $t=t_e$ to $t=t_a$), the compression threshold is multiplied by this value (default value: 1).\\
\texttt{select\_threshold\_ratio} & float & When using preselection for PT-MPO combination, the compression threshold is multiplied by this value (default value: 1). Note that \texttt{forward\_threshold\_ratio} or \texttt{backward\_threshold\_ratio} also apply depending on the sweep direction.\\
\texttt{final\_sweep\_n} & int & [argument] additional pairs of line sweeps are performed at the end of the PT-MPO generation (default value: 0)\\
\texttt{final\_sweep\_threshold} & float & Explicitly sets the threshold for final sweeps (default value: value of \texttt{threshold})\\
\texttt{add\_PT} & string [int] [int] & Read (read-only) PT-MPO from file (first argument = file name). Optionally, extend outer bond dimensions by a Hilbert space of dimension=[second argument] to the left and dimension=[third argument] to the right.\\
\texttt{initial\_PT} & string & Read and modify PT-MPO from file [argument]\\
\texttt{write\_PT} & string & Write generated or modified PT-MPO to file [argument]\\
\texttt{buffer\_blocksize} & int & Break up PT-MPO in blocks of size [argument] \\
\hline
\end{tabularx}
\caption{\label{tab:controls}Commands that can be specified in configuration files. For more commands for PT-MPO generation see Tab.~\ref{tab:generation}. The second column lists the type and order of arguments. A description is provided in the third column. Multiple arguments are separated by white spaces. Types are either text strings (string; must not contain whitespaces), floating point numbers (float), integers (int), matrix-valued expressions (matrix) as described in Sec.~\ref{sec:matrix-valued}, or boolean values (bool).
Floating point numbers can also be specified as 1x1 matrix-valued expressions. Arguments in square brackets are optional.}
\end{table*}

\begin{table*}
\renewcommand{\arraystretch}{1.25}
\begin{tabularx}{\linewidth}{|l |l |X|}
\hline
Command & Arguments & Comments \\
\tabsubtitle{PT-MPO method selection:}
\texttt{use\_combine\_tree} & bool & Selects the tree-like contraction scheme for ACE in Ref.~\cite{combine_tree} (default value: \texttt{false})\\
\texttt{use\_Gaussian} & bool & Selects the algorithm by J{\o}rgensen and Pollock~\cite{JP} for the generalized spin-boson model. All Gaussian methods use parameters of the \texttt{Boson\_...} generator (default value: \texttt{false})  \\
\texttt{use\_Gaussian\_divide\_and\_conquer} & bool & Selects the divide-and-conquer scheme of Ref.~\cite{DnC} for the generalized spin-boson model. (default value: \texttt{false})  \\
\texttt{use\_Gaussian\_periodic} & bool & Selects the periodic PT-MPO scheme of Ref.~\cite{DnC} for the generalized spin-boson model. Note that the memory time \texttt{t\_mem} should be specified (default value: \texttt{false}) \\
\tabsubtitle{Environment mode specification:}
\texttt{add\_single\_mode} & matrix matrix & A single environment mode is specified. The first argument contains the environment Hamiltonian $H_E$ on the Hilbert space $\mathcal{H}_S \otimes \mathcal{H}_\textrm{mode}$. The second argument is the initial density matrix of the mode
on $\mathcal{H}_\textrm{mode}$. \\
\texttt{add\_single\_mode\_from\_file}  & string matrix & A single environment mode is added, where the environment mode propagator is specified in the file [first argument].  The file has the format of a usual configuration file, where \texttt{add\_Hamiltonian}, \texttt{add\_Pulse}, \texttt{add\_Lindblad}, and \texttt{apply\_Operator\_...} are interpreted as acting on the composite Hilbert space $\mathcal{H}_S \otimes \mathcal{H}_\textrm{mode}$. The second argument is the initial mode density matrix.\\
\tabsubtitle{Boson generator:}
\texttt{Boson\_N\_modes} & int & Number of modes used to discretize the bosonic continuum. Ignored if any of the Gaussian methods are used.\\
\texttt{Boson\_M} & int & Hilbert space dimension per mode. Ignored if any of the Gaussian methods are used.\\
\texttt{Boson\_SysOp} & matrix & System operator in the system-environment interaction. See discussion of Eq.~\eqref{eq:DefSysOp}.  \\
\texttt{Boson\_J\_from\_file} & string & Coupling constants are defined by discretizing a spectral density provided in file [argument]. See discussion of Fig.~\ref{fig:example_ACE}(c). \\
\texttt{Boson\_J\_type} & string [...] & Use a predefined spectral density. See discussion of Fig.~\ref{fig:phonon_assisted}(a).\\
\texttt{Boson\_g} & float & Coupling constant to all modes is set equal to [argument] (units: ps$^{-1}$). See discussion of Fig.~\ref{fig:example_ACE}.\\
\texttt{Boson\_rate} & float & Coupling constant to all modes is set by matching the Markovian rate to [argument] (units: ps$^{-1}$). See discussion of Fig.~\ref{fig:example_ACE}.\\
\texttt{Boson\_omega\_min} & float & Lower frequency limit of mode continuum (unit: ps$^{-1}$; default value=0).\\
\texttt{Boson\_omega\_max} & float & Upper frequency limit of mode continuum (unit: ps$^{-1}$; default value=0).\\
\texttt{Boson\_temperature} & float & Sets the initial state of the bath as a thermal state with temperature [first argument] (units: K; default value: 0).\\
\texttt{Boson\_subtract\_polaron\_shift} & bool & Absorbs the polaron shift into a redefinition of system energies (default value: \texttt{true}).\\
\tabsubtitle{Fermion generator:}
\texttt{Fermion\_...} & ... & Same as the corresponding commands for the Boson generator, except for the following commands.\\
\texttt{Fermion\_EFermi} & float & Initial Fermi energy (units: meV; default value: -10$^6$ meV) \\
\hline
\end{tabularx}
\caption{\label{tab:generation}More control commands for PT-MPO generation. Same format as Tab.~\ref{tab:controls}.}
\end{table*}

In Tab.~\ref{tab:controls}, we provide a summary of commands to control the propagation of the open quantum system, the system propagator, and the handling of PT-MPOs. Further commands for controlling PT-MPO generation are listed in Tab.~\ref{tab:generation}.

\input{code.bbl}

\end{document}

%% file: code.bbl
%